\documentclass[journal,final,onecolumn]{IEEEtran}

\pdfoutput=1

\usepackage{amsmath,graphicx,amsfonts,amssymb,graphicx,url,times,epstopdf,comment,booktabs}

\usepackage{subcaption}
\usepackage{algorithm}
\usepackage{algorithmic}

\newcommand*{\iind}{f}
\newcommand*{\imax}{F}

\newcommand*{\jind}{n}
\newcommand*{\jmax}{N}

\newcommand*{\qind}{p}
\newcommand*{\qmax}{P}

\newcommand*{\mind}{m}
\newcommand*{\mmax}{M}

\newcommand*{\tind}{t}

\newcommand*{\kind}{q}
\newcommand*{\kmax}{Q}

\newcommand*{\Gind}{k}
\newcommand*{\Gmax}{K}

\newcommand*{\oind}{d}
\newcommand*{\omax}{D}

\newcommand*{\xtime}{{x}_\mind (\tind)}
\newcommand*{\stime}{{s}_\qind}
\newcommand*{\htime}{{h}_{\qind \mind \tind} (\tau)}

\newcommand*{\xFD}{\mathbf{x}_{\iind\jind}}
\newcommand*{\xFDhat}{\mathbf{\hat{x}}_{\iind\jind}}

\newcommand*{\yFDhat}{\mathbf{{y}}_{\iind\jind, \qind}}
\newcommand*{\hFD}{\mathbf{h}_{\iind\jind, \qind}}
\newcommand*{\sFD}{s_{\iind\jind, \qind}}
\newcommand*{\sFDhat}{\hat{s}_{\iind\jind, \qind}}
\newcommand*{\yFDsingle}{y_{\iind\jind,\qind}}

\newcommand*{\Xcov}{\mathbf{X}_{\iind\jind}}
\newcommand*{\Xcovhat}{\mathbf{\hat{X}}_{\iind\jind}}

\newcommand*{\Wmat}{\mathbf{W}_{\iind\oind}}

\newcommand*{\Hmat}{\mathbf{H}_{\iind\jind,\qind}}

\newcommand*{\tik}{t_{\iind \kind}}
\newcommand*{\vkj}{v_{\kind \jind}}
\newcommand*{\zpoj}{z_{\jind \oind, \qind}}
\newcommand*{\zpojt}{z_{\jind \oind, \qind}}
\newcommand*{\zoj}{S_{\oind \jind}}

\newcommand*{\zth}{T}

\newcommand*{\bkp}{b_{\kind, \qind}}

\newcommand{\ie}{\emph{i.e.}}

\newcommand*{\Zmu}{\mu_{\jind, \Gind}}
\newcommand*{\Zmup}{\hat{\mu}_{\jind, \qind}}
\newcommand*{\Zsigmap}{\hat{\sigma}_{\jind, \qind}^2}
\newcommand*{\Zanno}{\hat{\mu}_{\jind, \qind}^{\mathrm{(ann.)}}}
\newcommand*{\Zsigma}{\sigma_{\jind, \Gind}}
\newcommand*{\Za}{a_{\jind, \Gind}}

\newcommand*{\state}{\mathbf{s}}
\newcommand*{\trans}{\mathbf{A}}
\newcommand*{\measmod}{\mathbf{B}}
\newcommand*{\measur}{\mathbf{m}}
\newcommand*{\procnoise}{\mathbf{q}}
\newcommand*{\measnoise}{\mathbf{r}}

\DeclareMathOperator{\atan2}{atan2}

\begin{document}

% Title.
% ------
\title{Separation of Moving Sound Sources Using Multichannel NMF and Acoustic Tracking}

\author{
Joonas~Nikunen,~\IEEEmembership{}
~Aleksandr~Diment,~\IEEEmembership{}
and~Tuomas~Virtanen,~\IEEEmembership{Senior Member,~IEEE}% <-this % stops a space
\thanks{%Copyright (c) 2013 IEEE. Personal use of this material is
%permitted. However, permission to use this material for any other
%purposes must be obtained from the IEEE by sending a request to
%pubs-permissions@ieee.org.
%
J. Nikunen, A. Diment and T. Virtanen are with the Department of Signal
Processing, Tampere University of Technology, Tampere, Finland, email: firstname.lastname@tut.fi}
\thanks{This research was supported by Nokia Technologies.}
}

\maketitle

%\IEEEoverridecommandlockouts
%\IEEEpubid{\begin{minipage}{\textwidth}\ \\[12pt] \centering
%  \copyright 2017 IEEE. Personal use is permitted, but republication/redistribution requires IEEE permission.\\
%  See http://www.ieee.org/publications standards/publications/rights/index.html for more information.
%\end{minipage}} 

\begin{abstract}
In this paper we propose a method for separation of moving sound sources. The method is based on first tracking the sources and then estimation of source spectrograms using multichannel non-negative matrix factorization (NMF) and extracting the sources from the mixture by single-channel Wiener filtering. %The novelty of the proposed algorithm involves introducing 
We propose a novel multichannel NMF model with time-varying mixing of the sources denoted by spatial covariance matrices (SCM) and provide update equations for optimizing model parameters minimizing squared Frobenius norm. The SCMs of the model are obtained based on estimated directions of arrival of tracked sources at each time frame. The evaluation is based on established objective separation criteria and using real recordings of two and three simultaneous moving sound sources. The compared methods include conventional beamforming and ideal ratio mask separation. The proposed method is shown to exceed the separation quality of other evaluated blind approaches according to all measured quantities. Additionally, we evaluate the method's susceptibility towards tracking errors by comparing the separation quality achieved using annotated ground truth source trajectories. %The results prove that proposed method is suitable for completely blind operation for separation of moving sound sources. 
\end{abstract}


\IEEEpeerreviewmaketitle

\section{Introduction}
\label{sec:intro}
%scene using a mobile device, embodying several microphones, 

Separation of sound sources with time-varying mixing properties, caused by the movement of the sources,
is a relevant research problem for enabling intelligent audio applications in realistic operation conditions. These applications include, for example, speech enhancement and separation for automatic speech recognition \cite{barker2015third} especially when using voice commanded smart devices from afar \cite{farfieldrec}. Another emerging application field includes immersive audio for augmented reality \cite{valimaki2015assisted} which requires modification of the observed sound scene for example by removing sound sources and replacing them with augmented content. Separation of non-speech sources can be also used to improve sound event detection in multi-source noisy environment \cite{heittolaevent}. Most existing works related to sound separation are assuming stationary sources, and not many blind methods have targeted the problem of moving sound sources despite its high relevance in realistic conditions.
  %targeted the separation of moving sound sources. %, while machine learning approaches can address the problem by learning the spectral representation of the sources regardless of their spatial position with respect to receiving microphones.

The problem of sound source separation either from single or multi-channel recordings has been tackled with various methods over the years. The methods maximizing statistical independence of non-Gaussian sources, such as the independent component analysis (ICA) have been used for unmixing the sources in frequency domain \cite{SmaragdisICA,NestaICAseparation}. %\cite{SmaragdisICA,SawadaICAdoa,NestaICAseparation}. 
The concept of binary clustering of time-frequency blocks based on inter-channel cues, namely the level and the time difference, has resulted into class of separation methods based on time-frequency masking \cite{DUET,mask1}. Use of binary masks requires assuming that sound sources occupy disjoint time-frequency blocks \cite{Wdisjoint}. %The binary mask in time-frequency domain has been considered as the goal of computational audio scene analysis (CASA) \cite{casa}. 
More recently, single-channel speech enhancement and separation has been performed with the aid of machine learning and specifically by using deep neural networks (DNNs) \cite{DNN1,DNN2} for predicting the time-frequency masks for separation. Combining prediction of source spectrogram using DNNs and 
 spatial information in the form of source covariance matrices for audio source separation has been proposed in \cite{DNNcovar,DNNcovar2}.

% Additionally, microphone arrays with large number of elements allows the source separation also by means of beamforming \cite{Negbeamforming}. 

Another machine learning tool for masking based source separation is the spectrogram factorization by non-negative matrix factorization (NMF) and non-negative matrix deconvolution (NMD) which both have been widely utilized for speech separation and enhancement \cite{NMF1,gemmeke}. The NMF model decomposes mixture magnitude spectrogram into spectral templates and their time-dependent activations. In the case of single channel mixtures the separation is achieved by learning of noise or speaker dependent spectral templates from isolated sources in a training stage. %The NMF-based methods are usually operating with and utilize  

The NMF model can be extended for multichannel mixtures by incorporating spatial covariance matrix (SCM) estimation for the NMF components as in \cite{ozerovNTFmulti,FullrankNMF,sawadaCNMFtaslp,nikunenCNMFtaslp,nikunenCNMFicassp}. The analysis and introduction of spatial properties for NMF components allows separation based on spatial information, \ie, NMF components with similar spatial properties are considered to originate from the same sound source. These models require operation with complex-valued data and hereafter we refer to these extensions as multichannel NMF.  %The benefit of using NMF along with spatial covariance matrix estimation is that the spatial parameters can be estimated for an entire NMF component at once, which alleviates the affect of noise in observed spatial evidence compared time-frequency block-wise estimation.

%The multichannel extensions of NMF known as complex-valued non-negative matrix factorization (CNMF) have been considered for sound source separation in \cite{,sawadaCNMFwaspaasawadaCNMFtaslp,nikunenCNMFtaslp,nikunenCNMFicassp}. 

Recording of a realistic auditory scene often consists of sound sources which are moving with respect to the recording device, and conventional separation approaches assuming time-invariant mixing are not suitable for such a task. However, moving sound sources can be considered stationary within a short time block where the mixing can be assumed to be time-invariant. Using the block-wise approach for separation of moving sound sources requires merging the separated sources across individually processed blocks. For example, in block-wise ICA \cite{mukai_bICA1,mukai_bICA2,malek_bICA3} this is done by propagating the mixing matrix from the previous block and thus slowly adapting the mixing and preserving the source ordering in consecutive blocks. A recent generalization of multichannel NMF model \cite{ozerovNTFmulti} for time-varying mixing and separation of moving sound sources was proposed in \cite{kounades}. The reported results are promising, however the proposed algorithm requires using other state-of-the art source separation method in a blind setting for initialization.

Alternatively, separation of moving sources can be achieved by tracking the spatial position or direction of arrival (DOA) of the sources and using spatial filtering (beamforming or separation mask) for extracting the signal originating from the estimated position or direction in each time instance. In \cite{DOAHMM} the problem of DOA tracking and separation mask estimation is formulated jointly, however in this paper we consider a two stage approach where the acoustic tracking is done first and the separation masks are estimated in a separate (offline) stage. Also the separation masks are binary in \cite{DOAHMM} which will lead to compromised subjective separation quality even if oracle masks are used. 

Acoustic localization with microphone arrays can be achieved by transforming the time-difference of arrival (TDOA) obtained using generalized cross-correlation (GCC) into source position estimates \cite{brandstein}. Methods for estimation of trajectories of moving sound sources are based on Kalman filtering and its non-linear extensions \cite{ekf1,ekf2} for estimating the underlying state (position of the sound source) from the TDOA measurements. Alternatively, sequential Monte Carlo methods, \ie, particle filtering \cite{pf1,pf2} have been applied for tracking the position of the source based on TDOA measurements. For even more difficult case of multiple target tracking with data association problem, a Rao-Blackwellised particle filtering (RBPF) was proposed in \cite{sarkka1,sarkka2} and applied for acoustic tracking of multiple speakers in \cite{rbpf_aku1}. Additionally, the use of directional statistics and quantities wrapped on a unit circle or a sphere, such as the interchannel phase difference, has been recently considered for speaker tracking \cite{traaWrapped,traaTASLP}. 

%The difficulties in separation of moving sound sources include insufficient rejection of unwanted directions especially in the case of beamforming using only few microphones. In model-based separation, the problems include solving data association between adjacent input frames and having insufficient redundancy of spatial information for estimation of source or noise covariances since the time-varying mixing prevents excess integration over time. The data association can be solved by acoustic tracking of sources which in turn requires using separation model parameterizing sources based on their direction of arrival or spatial position.

\begin{figure*}[!ht]
  \centering
  \includegraphics[width = 1.0\linewidth]{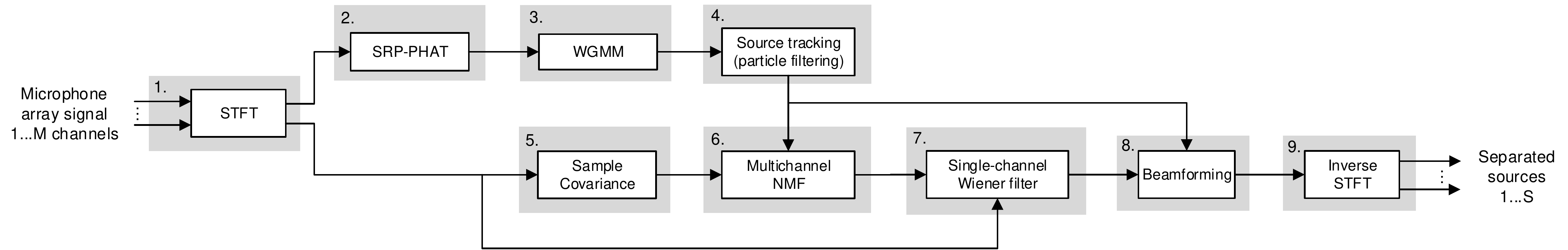}
	\caption{The block diagram of the proposed processing consisting of source tracking and multichannel NMF for separation of detected and tracked sound sources.}
  \label{fig:lohko}
	\vspace{-5mm}
\end{figure*}

In this paper we propose a separation method for moving sound sources based on acoustic tracking and estimation of source spectrogram from the tracked directions using the multichannel NMF with time-varying SCM model. %The estimated source spectrograms can be used to separate the sources from the mixture by means of Wiener filtering. 
The main contributions of this paper include: formulation of multichannel NMF model for time-varying mixing (moving sound sources),  integration of the spatial model with acoustic tracking to define spatial properties of sources in form of SCMs in each time frame and finally presenting the update equations for optimizing the multichannel NMF model parameters minimizing squared Frobenius norm. The parametrization of source DOA with directional statistics and using the tracker uncertainty for defining the SCMs of sources is a novel approach for representing the spatial location and spread of sound sources in the multichannel NMF. The acoustic tracker realization is combination of existing works on wrapped Gaussian mixture models \cite{WGMM} and particle filtering \cite{hartikainen}, but it will be presented in detail due to its output statistics are utilized in the proposed time-varying SCM model. % and multichannel NMF.

The evaluation of the proposed separation algorithm is based on objective separation criteria \cite{sisec,STOI,fwSNR} %, short time-objective intelligibility (STOI) \cite{STOI} and frequency weighted segmental SNR (fwSegSNR) \cite{fwSNR} 
and testing with mixtures of two and three simultaneous moving speakers. %The evaluation data includes movement of one or both of the speech sources. 
Additionally, hand-annotated source DOA trajectories are used for evaluating the performance of the acoustic tracker realization and studying the susceptibility of the proposed separation algorithm towards tracking errors. %by calculating its mean absolute error against annotations, additionally overall recall rate (source detection) is provided. 
%Annotated source trajectories are also used for studying the susceptibility of the separation mask estimation towards tracking errors by comparison to separation performance with annotated trajectories. 
The compared separation methods include conventional beamforming (DSB and MVDR) and upper reference is obtained by ideal ratio mask (IRM) separation \cite{IRM}. The proposed method achieves superior separation performance and the use of annotated trajectories shows no significant increase in separation performance, proving the proposed method suitable for realistic operation in a blind setting.

%The proposed method achieves superior separation performance  compared to beaforming and the acoustic tracking realization achieves recall rate of 82\% with mean absolute error less than $10$ azimuthal degrees. % Additionally, using the annotated trajectories shows no significant increase in separation performance, proving the proposed method suitable for realistic operation in a blind setting.

The paper is organized as follows. First the problem of separating moving sound sources and an overview of the proposed processing is given in Section \ref{sec:overview}. Next we introduce directional statistics and describe the acoustic source tracker realization in Section \ref{sec:tracking}. In Section \ref{sec:CNMF} the multichannel NMF separation model for sources with time-varying mixing is proposed and the utilization of tracker output within the separation model is explained. In Section \ref{sec:Eval} the tracking and separation performance of the proposed algorithm is evaluated using real recorded test material captured using a compact four-element microphone array. The work is concluded in Section \ref{sec:Conclusion}.

\section{Problem statement and algorithm overview}
\label{sec:overview}

\subsection{Mixing model of moving sound sources}
A microphone array composed of microphones $(\mind = 1, \dots, \mmax)$ observes a mixture of $\qind = 1, \dots, \qmax$ source signals $\stime (\tind)$ sampled at discrete time instances indexed by $\tind$. The sources are moving and have time-varying mixing properties, denoted by room impulse response (RIR) $\htime$, for each time index $\tind$. The resulting mixture signal can be given as
\begin{equation}
\label{eq:timemodel}
\xtime = \sum_{\qind=1}^\qmax \sum_{\tau} \stime (\tind - \tau) \htime. %\quad \tind' = \tind-\tau.
\end{equation}
In sound source separation the aim is to estimate the source signals $\stime$ and their mixing $\htime$ by only observing $\xtime$.

%\subsection{Mixture in time-frequency domain}
In this paper audio is processed in frequency domain obtained using the short time Fourier transform (STFT). The STFT of a time-domain mixture signal is calculated by dividing the signal into short overlapping frames, applying a window function and taking the discrete Fourier transform (DFT) of the windowed frame. The mixing properties denoted by the time-dependent RIRs $\htime$ change slowly over time and in practice the difference between adjacent time indices $\tind$ is small, thus we can consider mixing being constant within a small time window. This allows to approximate the time-dependent mixing \eqref{eq:timemodel} in time-frequency (TF) domain as
\begin{equation}
\label{eq:FDmodel}
\xFD \approx \sum_{\qind=1}^\qmax \hFD \sFD =  \sum_{\qind=1}^\qmax \yFDhat.
\end{equation}
The STFT of the mixture signal is denoted by $\xFD = [x_{\iind\jind 1},\dots,x_{\iind\jind \mmax}]^T $ for each TF-point $(\iind,\jind)$ of each input channel $(\mind = 1, \dots, \mmax)$. The single-channel STFT of each source $\qind$ is denoted by $\sFD$ and their frequency domain RIRs (fixed withing each time frame $\jind$) are denoted by $\hFD = [h_{\iind\jind 1},\dots,h_{\iind\jind \mmax}]^T$. The source signals convolved with the impulse responses are denoted by $\yFDhat$.

\subsection{Overview of the processing}
The proposed method consists of source spectrogram estimation based on the DOA of the sources of interest in each time frame and the estimated spectrograms are used for separation mask generation by generalized Wiener filter. The processing is based on two stages: the acoustic tracker and the offline separation mask estimation by multichannel NMF. The block diagram of the method is illustrated in Figure \ref{fig:lohko}. The source tracking branch operates frame-by-frame and can be though as online algorithm while the parameters of the multichannel NMF model are estimated from the entire signal at once (offline). % and thus the overall algorithm is explained in a stepwise manner. %We will not deal with the block-wise NMF processing in this paper although it is a straightforward extension and can handle processing of continuous signals.

First the STFT of the input signals is calculated. The tracking branch starts with calculating the steered response power (SRP) of the signal under analysis. SRP denotes the spatial energy as a function of DOA for each time frame. A wrapped Gaussian mixture model (WGMM) \cite{WGMM} of the SRP function in each time frame is estimated, which converts spatial energy histogram (\ie~the SRP) into DOA measurements. WGMM parameters are used as measurements in acoustic tracking which is implemented using particle filtering \cite{hartikainen}. The multi-target tracker detects the births and deaths of sources, solves the data-associations of measurement belonging to one of existing sources and predicts the source trajectories. % using either Kalman filter (KF) or extended Kalman filter (EKF).

%and estimating a NMF-based spectral model of tracked sources to separate them by time-frequency filtering. %First an initial CNMF model is estimated which parameterizes the spatial location of the sources using a DOA-based spatial covariance matrix (SCM) model \cite{nikunenCNMFtaslp}. 
%The so called direction weights of the SCM model denote the source location in terms of discrete DOA distribution at each given time frame. 

In the second stage a spatial covariance matrix model (SCM model) \cite{nikunenCNMFtaslp} parameterized by DOA is defined based on the acoustic tracker output (source DOA at each time-frame). The obtained SCMs denote the spatial behavior of sources over time and a spectral model of sources originating from the tracked direction is estimated using multichannel NMF. The multichannel source signals are reconstructed using a single-channel Wiener filter based on the estimated spectrogram of each source and single-channel signals are obtained by applying the delay-and-sum beamforming to the separated multichannel signals. Finally, time-domain signals are reconstructed by applying inverse STFT and overlap-add.

\section{Source trajectory estimation}
\label{sec:tracking}
The goal of the first part of the proposed algorithm is to estimate DOA trajectories of the sound sources that are to be separated. The process consist of three consecutive steps: calculating the spatial energy emitted from all directions (Section \ref{sec:TDOA}), converting the discrete spatial distribution into DOA measurements (Sections \ref{sec:WGMM} and \ref{sec:DOAmes}) and multi-target tracking consisting of source detection, data-association and source trajectory estimation (Section \ref{sec:raoPF}).

\subsection{Time-difference of arrival and steered response power}
\label{sec:TDOA}
Spatial signal processing with spaced microphone arrays is based on observing time delays between the array elements. In far-field propagation the wavefront direction of arrival corresponds to a set of TDOA values between each microphone pair. We start by defining a unit direction vector $\mathbf{k} \in \mathcal{R}^3, ||\mathbf{k}|| = 1$ originating from the geometric center of the array $\mathbf{p} = [0,0,0]^T$ and pointing towards direction %indexed by $\oind$. The direction is 
parametrized by azimuth $\theta \in [0, 2\pi]$ and elevation $\varphi \in [0, \pi]$. Given a microphone array consisting of two microphones  $\mind_1$ and $\mind_2$  at locations $\mathbf{\mind}_1 \in \mathcal{R}^3$, $\mathbf{\mind}_2 \in \mathcal{R}^3$ the TDOA between them for a sound source at direction $\mathbf{k}$ is obtained as 
\begin{equation}
\label{eq:TDoA}
\tau(\mind_1,\mind_2) = \frac{ -\mathbf{k}^T (\mathbf{m}_1 - \mathbf{m}_2) } {v},
\end{equation}
where $v$ is the speed of sound. The above TDOA corresponds to a phase difference of $\exp (- j \omega_\iind \tau(\mind_1,\mind_2))$ in the frequency domain, where $\omega_\iind = 2 \pi  (\iind-1) F_s/{N}$ ($F_s$ is the sampling frequency and $N$ is the STFT window length). 
%($\fHz_\iind = (\iind-1) F_s/{N}$ is frequency in Hz of $\iind$th STFT bin, $F_s$ is the sampling frequency and $N$ is the STFT window length).

%Any given array geometry can be translated and rotated in such a way that its geometrical center is located at the origin of the coordinate system, which is done to simplify the TDOA calculations. 

From now on we operate with a set of different directions indexed by $\oind = 1,\dots,\omax$ and the direction vector corresponding to $\oind$th direction is defined as $\mathbf{k}_\oind$ resulting to TDOA of $\tau_\oind(\mind_1,\mind_2)$. The spatial energy originating from the direction $[\theta_\oind, \varphi_\oind]$ at each time frame $\jind$ can be calculated using the steered response power (SRP) with PHAT weighting \cite{tashevDOAkirja} defined as
\begin{equation}
\label{eq:SRP}
\zoj = \sum_{\mind_1=1}^{\mmax-1} \sum_{\mind_2=\mind_1+1}^\mmax \sum_{\iind=1}^\imax \frac{x_{\iind\jind\mind_1} x_{\iind\jind\mind_2}^*} {|x_{\iind\jind\mind_1} x_{\iind\jind\mind_2}^*|} \exp (j \omega_\iind \tau_\oind(\mind_1,\mind_2)),
\end{equation}
where ${ }^*$ denotes complex-conjugate and the term  $\exp (j \omega_\iind \tau_\oind(\mind_1,\mind_2))$ is responsible for time-aligning the microphone signals. SRP denotes the spatial distribution of the mixture consisting of spatial evidence from multiple sources and searching for multiple local maxima of the SRP function at a single time frame $\jind$ corresponds to DOA estimation of sources present in that time frame. Repeating the pick peaking for all time frames of SRP would result to DOA measurements that are permuted over time and subsequently in the tracking stage the permuted DOA measurements are associated to multiple sources over time.

\begin{figure}[tb]
  \centering
  \includegraphics[width = 0.5\linewidth]{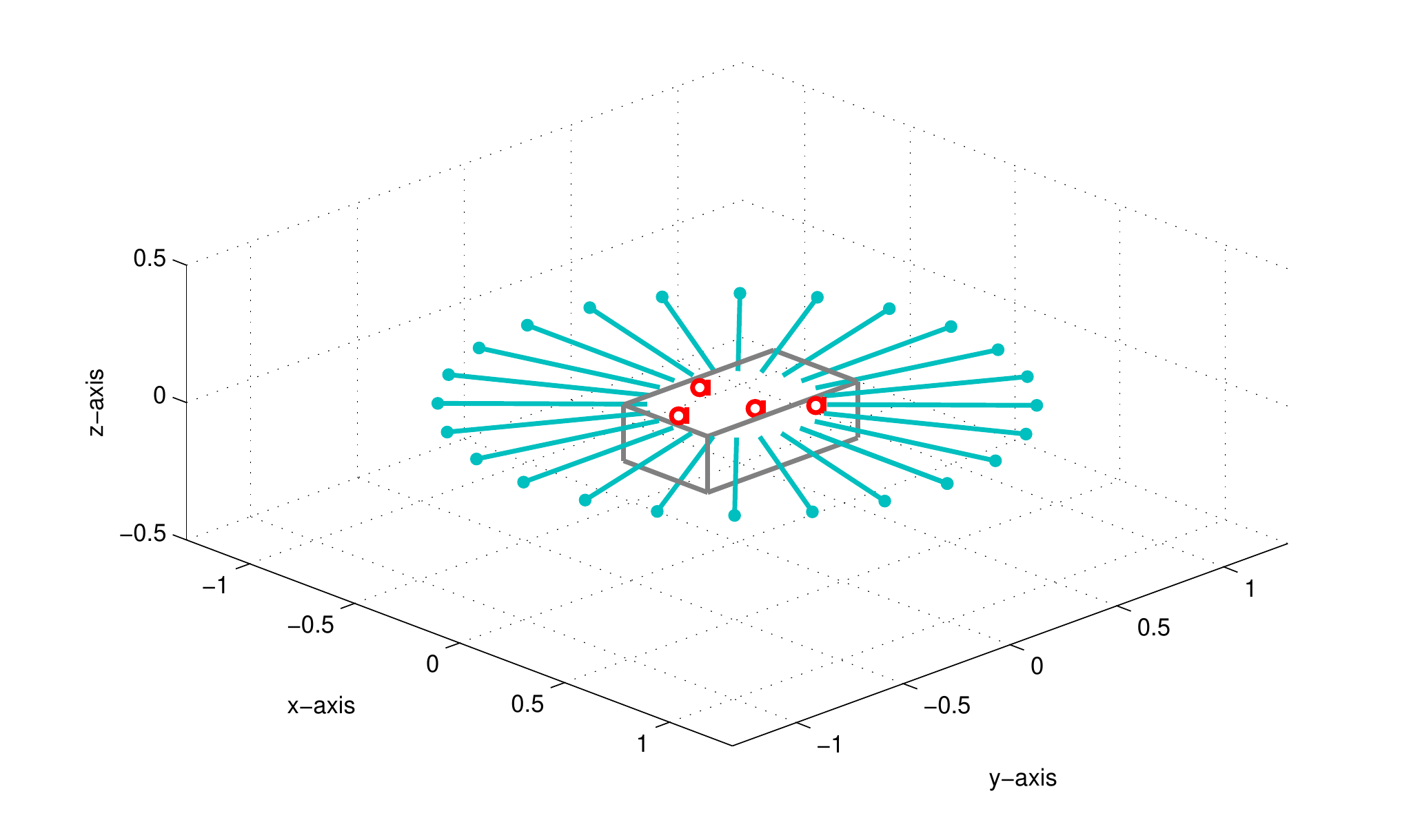}
 \caption{Illustration of a sparse grid of direction vectors by lines with dot in end and an example array enclosure and enclosed microphones (circles).}
\label{fig:leikkaus}
\vspace{-4mm}
\end{figure}

In a general case the directions $\oind = 1,\dots,\omax$ would uniformly sample a unit sphere around the array, but in this paper we only consider the zero elevation plane, \ie, $\varphi_\oind = 0 \; \forall \oind$. We assume that the sources of interest lie approximately on the xy-plane with respect to the microphone array and the directional statistics used in tracking of the sources simplifies to a univariate case. % when the sampling of the spatial space around the array is by azimuthal information only.
A sparse grid of directions vectors with spacing of adjacent azimuths by $\frac{\pi}{12}$ is illustrated in Figure \ref{fig:leikkaus} along with the array casing and microphones corresponding to the actual compact array used in the evaluations.

\subsection{Wrapped Gaussian mixture model}
\label{sec:WGMM}
Instead of searching peaks from the SRP \eqref{eq:SRP}, we propose to model the mixture spatial distribution using a wrapped Gaussian mixture model (WGMM) estimated separately for each time-frame of the SRP. The estimation of the parameters of the WGMM results to converting the discrete spatial distribution obtained by SRP into multiple DOA measurements with mean, variance and weight. The individual wrapped Gaussians model the spatial evidence caused by the actual sources while some of them may model the noise or phantom peaks in the SRP caused by sound reflecting from boundaries. The use of WGMM alleviates effect of noise when the width of each peak, denoted by variance of each wrapped Gaussian, can be used to denote measurement uncertainty in the acoustic tracking stage. 

The probability density function (PDF) of univariate wrapped Gaussian distribution \cite{WGMM,SmaragdisWG,TraaMscThesis} with mean $\mu$ and variance $\sigma^2$ can be defined as
\begin{align}
P(\theta; \mu, \sigma^2) = & \sum_{l=-\infty}^{\infty} \mathcal{N}(\theta; \mu + l 2 \pi,\sigma^2) \nonumber \\
 = & \sum_{l=-\infty}^{\infty} \frac{1}{\sqrt{2 \pi \sigma^2}} e^{- \frac{(\theta - \mu + 2 \pi l)^2}{2\sigma^2 }},
\end{align}
where $\mathcal{N}(\theta; \mu,\sigma^2)$ is a PDF of a regular Gaussian distribution, $l$ is the wrapping index of $2\pi$ multiples and $\theta \in [-\pi, \pi]$. The multivariate version of the wrapped Gaussian distribution is given in \cite{WGMM}, but it is not of interest in this paper.

The WGMM with weights $a_\Gind$ for each wrapped Gaussian distribution $\Gind$ is defined as
\begin{equation}
\label{eq:WGMMpdf}
P (\theta; \mathbf{a}, \boldsymbol\mu, \boldsymbol\sigma^2 ) = \sum_{\Gind=1}^\Gmax a_\Gind \sum_{l=-\infty}^{\infty} \mathcal{N}(\theta; \mu_\Gind + l 2 \pi,\sigma_\Gind^2).
\end{equation}
where $\Gmax$ is the total number of wrapped Gaussians in the model and EM algorithm for estimating parameters $\{\mathbf{a}, \boldsymbol\mu, \boldsymbol\sigma^2\}$ that maximize the log likelihood 
\begin{equation}
\label{eq:WGMMlogL}
\log L = \sum_{\oind=1}^\omax \log \sum_{\Gind=1}^\Gmax a_\Gind \sum_{l=-\infty}^{\infty} \mathcal{N}(\theta_\oind; \mu_\Gind + l 2 \pi,\sigma_\Gind^2),
\end{equation}
is given in \cite{WGMM,TraaMscThesis}. The parameter $\theta_\oind$ denotes the azimuth angles of the directions indices $\oind = 1,\dots,\omax$ used to calculate SRP in Equation \eqref{eq:SRP}.

\begin{figure}[tb]
  \centering
  \includegraphics[width = 0.5\linewidth]{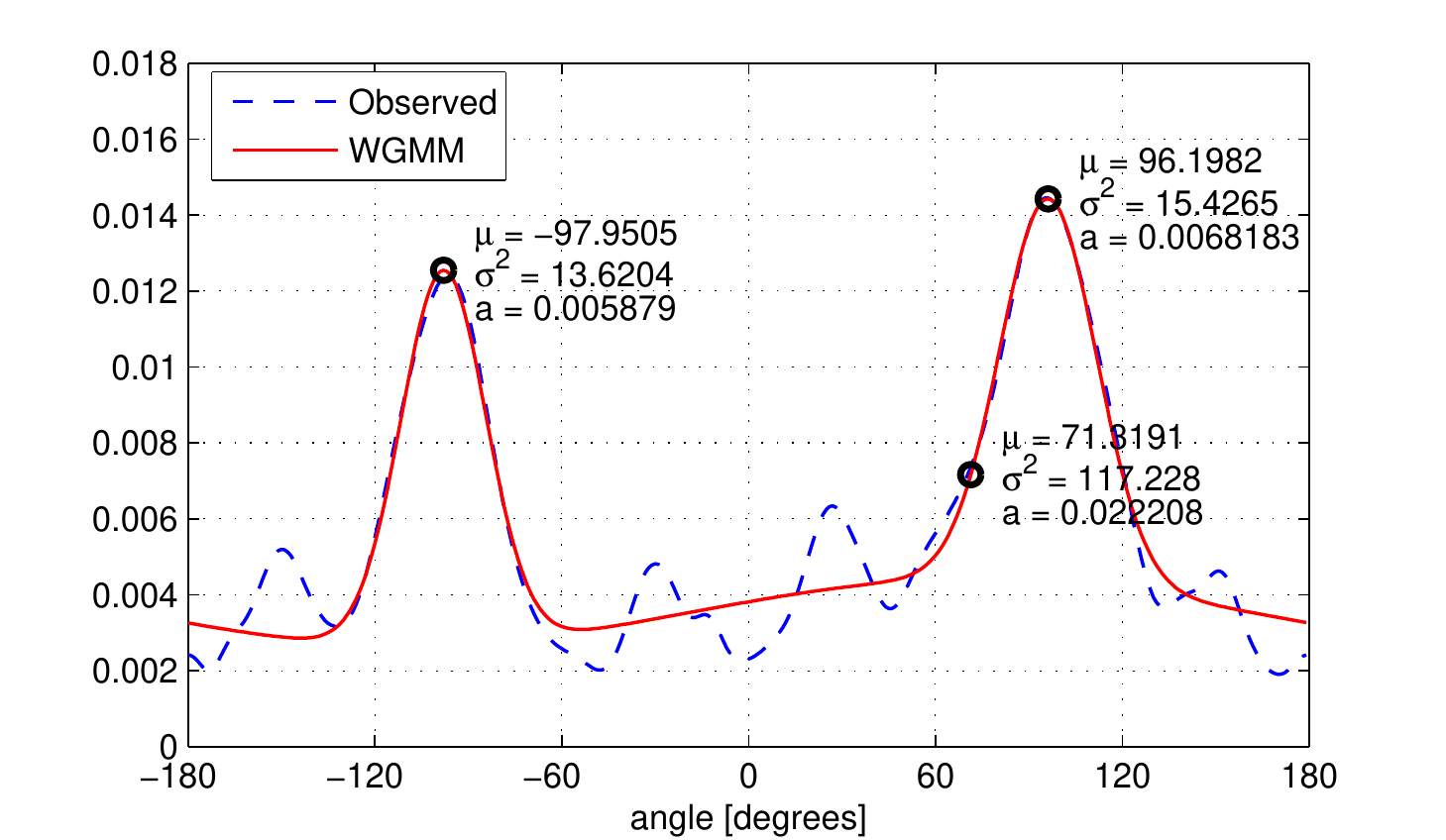}
	\caption{The observed SRP $\zoj$ for a single time-frame $\jind$ and the WGMM with $\Gmax = 3$ estimated from it.}
  \label{fig:WGMM}
\end{figure}

%In the case of the proposed SCM model in Equation \eqref{eq:SCMmodel} the observed data, the direction weights $\zpoj$, is a discrete/sampled estimate of the continuous DOA function of each source. Similarly, 
The EM-algorithm for WGMM as presented in \cite{WGMM,TraaMscThesis} requires observing data points generated by the underlying distribution whereas $\zoj$ for a single frame $\jind$ is effectively a histogram denoting spatial energy emitted from each scanned direction indexed by $\oind$. % whereas $\zoj$ correspond to observing the discretely sampled mixture distribution, \ie, the histogram of the data. 
Estimating the WGMM parameters based on the histogram requires modification of the algorithm presented in \cite{WGMM,TraaMscThesis} to account for inputs consisting of discretely sampled mixture distribution, \ie, the histogram bin values. %WGMM parameter estimation by EM-algorithm to account data points with different weights. 
The modification results to Algorithm \ref{alg:EMWGMM} where SRP of a single frame $\jind$ is denoted by $s_\oind$ and the updates are iterated until converge of $\eta_{\oind \Gind l}$. Prior knowledge can be used to set initial values for $a_\Gind, \mu_\Gind$ and $\sigma_\Gind^2$ or they can be initialized randomly. An example of the SRP $\zoj$ of a single time-frame and three component WGMM estimated from it is illustrated in Figure \ref{fig:WGMM}.

\renewcommand{\algorithmicrequire}{\textbf{Input:}}
\renewcommand{\algorithmicensure}{\textbf{Output:}}

\begin{algorithm}[tb]
\caption{EM-algorithm for estimation of WGMM model parameters for a histogram $s_\oind$ (single frame from the entire SRP $\zoj$).}
\label{alg:EMWGMM}
\begin{algorithmic}
\REQUIRE Histogram data $s_\oind$ and initial values for $a_\Gind, \mu_\Gind$ and $\sigma_\Gind^2$
\STATE \textbf{E-STEP}
\STATE $\eta_{\oind \Gind l} = \frac{\mathcal{N}(\theta_\oind; \mu_\Gind + l 2 \pi,\sigma_\Gind^2) a_\Gind} {\sum_{\Gind = 1}^{\Gmax} \sum_{l=-\infty}^{\infty} \mathcal{N}(\theta_\oind; \mu_\Gind + l 2 \pi,\sigma_\Gind^2) a_\Gind } $
\STATE \textbf{M-STEP}
\STATE $\mu_\Gind = \frac{ \sum_{\oind=1}^{\omax} \sum_{l=-\infty}^{\infty} (\theta_\oind - 2 \pi l) \eta_{\oind \Gind l} s_\oind } {\sum_{\oind=1}^{\omax} \sum_{l=-\infty}^{\infty}  \eta_{\oind \Gind l} s_\oind }$
\STATE $\sigma_\Gind^2 = \frac{ \sum_{\oind=1}^{\omax} \sum_{l=-\infty}^{\infty} (\theta_\oind - \mu_\Gind - 2 \pi l) \eta_{\oind \Gind l} s_\oind } {\sum_{\oind=1}^{\omax} \sum_{l=-\infty}^{\infty}  \eta_{\oind \Gind l} s_\oind }$
\STATE $a_\Gind = \frac{1}{\sum_{\oind} s_\oind} \sum_{\oind=1}^{\omax} \sum_{l=-\infty}^{\infty} \eta_{\oind \Gind l} s_\oind $
\end{algorithmic}
\end{algorithm}

\subsection{DOA measurements by WGMM}
\label{sec:DOAmes}
Algorithm \ref{alg:EMWGMM} is applied individually for each time frame $\jind = 1,\dots,\jmax$ of $\zoj$. A mixture of $\Gind = 1, \dots, \Gmax$ wrapped Gaussians for each time frame is obtained and the resulting means $\Zmu$ with variances $\Zsigma$ and weights $\Za$ are considered as permuted DOA measurements. At this point of the algorithm it is unknown which of the measurements $\Gind = 1,\dots,\Gmax$ in each frame $\jind$ are caused by actual sources and which correspond to noise. Also the detection of sources and association of different measurements $\Gind$ to sources $\qind$ over time is unknown, i.e. $\Gind$th measurement in adjacent frames may be caused by different sources. The source detection, data-association and actual source trajectory estimation is solved using the Rao-Blackwellized particle filtering introduced in Section \ref{sec:raoPF}.

Random initialization of Algorithm \ref{alg:EMWGMM} could be used in each time frame. However, initial values close to the optimal ones speed up the algorithm convergence and in practice estimates from previous frame can be used as initialization for the subsequent frame. Please note that this initialization strategy does not guarantee preserving any association between $\Gind$th wrapped Gaussians in adjacent frames and ordering need to be considered as permuted.

%The WGMM parameter estimation in 1-D case with observed discretized distribution can be also formulated as a following minimization problem,
%\begin{equation}
%\label{eq:nonlinLS}
%\{\Zmu, \Zsigma, \Za\} \leftarrow \min_{ \mathbf{a}, \boldsymbol\mu, \boldsymbol\sigma^2} \sum_{\oind=1}^{\omax} (\zoj - P(\theta_\oind; \mathbf{a}_\jind, \boldsymbol\mu_\jind, \boldsymbol\sigma_\jind^2))^2.
%\end{equation}
%where the $\theta_\oind$ corresponds to the azimuthal angles of the direction vectors used in calculation of the SRP. The above minimization is a conventional non-linear least squares problem and numerous methods for solving its parameters exist in literature \cite{trustregion}, however these were found computationally quite unfeasible. 

The WGMM parameters $\{\Zmu, \Zsigma, \Za\}$ are hereafter referred to as DOA measurements regarding the acoustic tracking: $\Zmu$ are the measurement means, $\Zsigma$ denote measurement reliability and $\Za$ are the proportional weights of the measurements. Given that a WGMM with $\Gmax$ components for each time frame is estimated, not all WGMM components are caused by actual spatial evidence but are merely modeling the noise floor and phantom peaks in the SRP. The situation is illustrated in Figure \ref{fig:WGMM}, where the third WGMM component with mean $\mu = 71^\circ$ has a very high variance $\sigma^2 = 117^\circ$ in comparison to the actual observable peaks. 

By investigating the variance and weight of each WGMM component the false measurements can be efficiently removed before applying the actual tracking algorithm. The means $\Zmu$ for each time frame for an arbitrary test signal and after removing measurements with $\Zsigma %> 0.6 \; \mathrm{rad} \approx 
>  35^\circ$ or $\Za < 0.15$ are illustrated in Figure \ref{fig:WGMMmeans}. The removal of false measurements reveals two distinct observable trajectories, however the data association between each frame is unknown at this stage. The thresholds for measurement removal can be set to be global (signal and capturing environment independent) and their choice is discussed in more details in Section \ref{sec:Eval}.

\subsection{Acoustic tracking of multiple sound sources}
\label{sec:raoPF}
The problem setting in tracking of multiple sound sources is as follows. Multiple DOA measurements are obtained in each time frame and the task is to decide whether the new measurement is 1) associated to an existing source 2) identified as clutter, 3) evidence of new source (birth) and finally 4) determining possible deaths of existing sources. After the data-association step the dynamic state of the active sources is updated and particularly it is required to preserve the source statistics over short inactive segments (pauses between words in speech) by prediction based on previous state of the source (location, velocity and acceleration).

%\begin{figure}[tp]
  %
  %\includegraphics[width = 0.95\linewidth]{journal_WGMMallEM}
	%\caption{The WGMM means $\Zmu$ for each time-frame $\iind$.}
  %\label{fig:WGMMmeans}
%\end{figure}
%
%\begin{figure}[tp]
  %\centering
  %\includegraphics[width = 0.95\linewidth]{journal_WGMMcleanEM}
	%\caption{The WGMM means $\Zmu$ for each time-frame $\iind$ after removing measurements with $\Zsigma > 1 \; \mathrm{rad} \; (57^\circ)$ and $\Za < 0.005$ }
  %\label{fig:WGMMmeans2}
%\end{figure}

%\subsubsection{State Space Model}
In the following, we shortly review the Rao-Blackwellized particle filter (RBPF) framework proposed in \cite{sarkka1} for the problem of multi-target tracking and use its freely available implementation\footnote{\url{http://becs.aalto.fi/en/research/bayes/rbmcda/}} documented in \cite{hartikainen}. We give the state-space representation of the dynamical system that is being tracked but we do not go to any further details of RBPF. The algorithm proposed in \cite{sarkka1} and the associated implementation has been used in \cite{spille} for tracking the azimuth angle of speakers in a similar setting.

%The particle filter for multiple target tracking\footnote{Implementation freely available from \url{http://becs.aalto.fi/en/research/bayes/rbmcda/} and documented in \cite{hartikainen}} proposed in \cite{sarkka1} is used to solve the the data association, the target detection and tracking of source DOA trajectories underlying in the measurements produced by the WGMM model. 
% We give the state-space representation of the dynamical system that is being tracked but we do not go to any further details of particle filtering.

%In RBPF the problem multi-target tracking can be conceptually divided into data association (assigning a measurement to existing target) and tracking of single targets (dynamic model for targets). 

Multi-target tracking by RBPF is essentially based on dividing the entire problem into two parts, estimation of data association and tracking of single targets. This can be done with the Rao-Blackwellization procedure \cite{sarkka2} where the estimation of the posterior distribution of the data associations is done first and then applying single target tracking sub-problem conditioned on the data associations. Adding the estimation of unknown number of targets \cite{sarkka1} using a probabilistic model results into RBPF framework that solves the entire problem of tracking unknown number of targets. The benefit of the Rao-Blackwellization is that by conditioning the data association allows calculating the filtering equations in closed form instead of using particle filtering and data sampling based techniques for all steps, which leads to generally better results.

%In RBPF the problem is divided into tracking of single targets (dynamic model for each target in each particle), data association (assigning a measurement to existing target) and estimating the number of targets (if measurements evokes a new target or one of the targets disappears according to the target life time probability density). 

%First, the posterior distribution of the data association is calculated using a sequential-importance resampling SIR, particle-filtering algorithm. Second, the single targets are tracked by an extended Kalman filter that depends on the data associations. Rao-Blackwellization exploits the fact that it is often possible to calculate the filtering equations in closed form.

\subsubsection{State-space model for speaker DOA tracking}
In the RBPF framework the single target tracking consist of Bayesian filtering which requires defining the dynamic model and the measurement model of the problem. For the time being we omit the WGMM component index $\Gind$ and the source index $\qind$ and define the state space model equations for the single target tracking sub-problem. 

The goal is to estimate the state of the dynamical system in each time instance $\jind$ and the state in our case is defined as a 2-D point $(x,y)$ at the unit circle with velocities along both axes $(\dot{x},\dot{y})$ defined as
\begin{align}
\label{eq:statevec}
\state_\jind & = \begin{bmatrix} x_\jind , & y_\jind , & \dot{x}_\jind , & \dot{y}_\jind , \end{bmatrix}^T.
\end{align}
The angle of the x-y coordinate $(x_\jind , y_\jind)$ represents the DOA of the source and it avoids dealing with the $2\pi$ ambiguity of \mbox{1-D} DOA variables in the dynamic model. %In practice the state vector is defined individually for each detected source $\qind = 1,\dots,\qmax$ and is denoted hereafter as $\state_\jind^{(\qind)}$.
%In practice the state vector is defined individually for each detected source $\qind = 1,\dots,\qmax$ and is denoted hereafter as $\state_\jind^{(\qind)}$.
%The discrete time linear-Gaussian state space model consist of the dynamic model and measurement model. 
The dynamic model that predicts the target state based on previous time step is defined as 
\begin{align}
\state_\jind = & \trans_{\jind-1} \state_{\jind-1} + \procnoise_{\jind-1}
\end{align}
where $\trans_{\jind-1}$ is the state transition matrix and $\procnoise_{\jind-1} \sim \mathcal{N}(0, \lambda^2 \mathbf{I})$ is the process noise. With the above definition for state $\state_\jind$ the transition matrix becomes linear and is defined as,
\begin{align}
\label{eq:TransM}
\trans_{\jind-1} = \begin{bmatrix} 
1 & 0 & \Delta t & 0 \\ 
0 & 1 & 0 & \Delta t \\  
0 & 0 & 1 & 0 \\ 
0 & 0 & 0 & 1 \end{bmatrix},
\end{align}
where $\Delta t$ is the time difference between consecutive time steps. The resulting dynamic model can be described as follows: the predicted DOA at current time step $\jind$ is the DOA of the previous time step in x-y coordinates added with its velocity in previous time step multiplied with the time constant, \ie, the time between consecutive processing frames.

For the measurement representation we use the rotating vector model \cite{RotVecMod} that converts the wrapped 1-D angle measurements $\mu \in [0,2\pi]$ to a 2-D point on a unit circle, resulting in measurement vector %$\measur_\jind \in \mathbb{R}^{2}$ defined as
\begin{align}
\label{eq:measvec}
\measur_\jind = \begin{bmatrix} \cos(\mu) , \sin(\mu) \end{bmatrix}^T.
\end{align}
The measurement model is defined as,
\begin{align}
\measur_\jind = & \measmod_{\jind} \state_{\jind} + \measnoise_{\jind},
\end{align}
where $\measmod_{\jind}$ is the measurement model matrix and $\measnoise_{\jind} \sim \mathcal{N}(0, \sigma \mathbf{I})$ is the measurement noise. The measurement model matrix $\measmod_{\jind}$ converts the state $\state_{\jind}$ into measurement $\measur_\jind$ (x-y coordinates) simply by omitting the velocities and is defined as,
\begin{align}
\label{eq:MeasM}
\measmod_{\jind}= \begin{bmatrix} 
1 & 0 & 0 & 0 \\ 
0 & 1 & 0 & 0 \end{bmatrix}.
\end{align}
The above definitions result to linear dynamic and and measurement model matrices \eqref{eq:TransM} and \eqref{eq:MeasM} allowing use of regular Kalman filter equations to update and predict the state of the particles in RBPF framework \cite{hartikainen}.

%For the measurement representation we use the rotating vector model \cite{RotVecMod} that converts the wrapped 1-D angle measurements $\Zmu$ to a 2-D point at unit circle, resulting in measurement vector $\measur_\jind^{(\Gind)} \in \mathbb{R}^{2}$ defined as
%\begin{align}
%\label{eq:measvec}
%\measur_\jind^{(\Gind)} = \begin{bmatrix} \cos(\Zmu) \\ \sin(\Zmu) \end{bmatrix},
%\end{align}
%The second part of the state space model is the measurement model defined as,
%\begin{align}
%\measur_\jind = & \measmod_{\jind} \state_{\jind}+ \measnoise_{\jind},
%\end{align}
%where $\measmod_{\jind}$ is the measurement model matrix and $\measnoise_{\jind}^{(\Gind)} \sim \mathcal{N}(0, \Zsigma \mathbf{I})$ is the measurement noise. 

%($\state_\jind \in \mathbb{R}^{4}$), %Similarly, the state is defined as a 2-D point at the unit circle with constant velocity model along both axis ($\state_\jind \in \mathbb{R}^{4}$), resulting to
%\begin{align}
%\label{eq:statevec}
%\state_\jind^{(\qind)} = \begin{bmatrix} x \\ y \\ \dot{x} \\ \dot{y} \end{bmatrix},
%\end{align}

%where the superscript ${ }^{(\qind)}$ denotes multiple detected sources. % $\dot{x}$ and $\dot{y}$ are velocities along x and y-axis, respectively. 

%, and for linear systems with linear dynamics the well known optimal filtering by Kalman filter \cite{kalman} exists. 

We acknowledge that the dynamic system used here is theoretically imperfect with respect to using 2-D quantities while the state and the measurements are truly 1-D, leading to additional noise in the system as pointed out in \cite{traaWrapped}. However, during the implementation of the acoustic tracker the chosen linear models were found performing better than the non-linear alternatives. Alternatively, the problem of tracking wrapped quantities using 1-D state could be addressed via wrapped Kalman filtering as proposed in \cite{traaWrapped}.

%\subsubsection{Measurement model}

%In order to make the measurement model matrix and state transition linear we use the rotating vector model \cite{RotVecMod} that converts the wrapped 1-D angle measurements to a 2-D point at unit circle. 

%\begin{align}
%\state_\jind^{(\qind)} & = \begin{bmatrix} x & y & \dot{x} & \dot{y} \end{bmatrix}^T \\ 
%\measur_\jind^{(\Gind)} & = \begin{bmatrix} \cos(\Zmu) & \sin(\Zmu) \end{bmatrix}^T.
%\end{align}

%The Bayesian filtering of a single particle state requires definition of the state-space model. 

%\subsubsection{Bayesian Filtering of Wrapped Data}
%During the development and implementation of acoustic tracking it was found out that using measurement model matix (state -> measurement)

\begin{figure}[tb]
  \begin{subfigure}[t]{0.5\linewidth}
  \includegraphics[width = 1.0\linewidth]{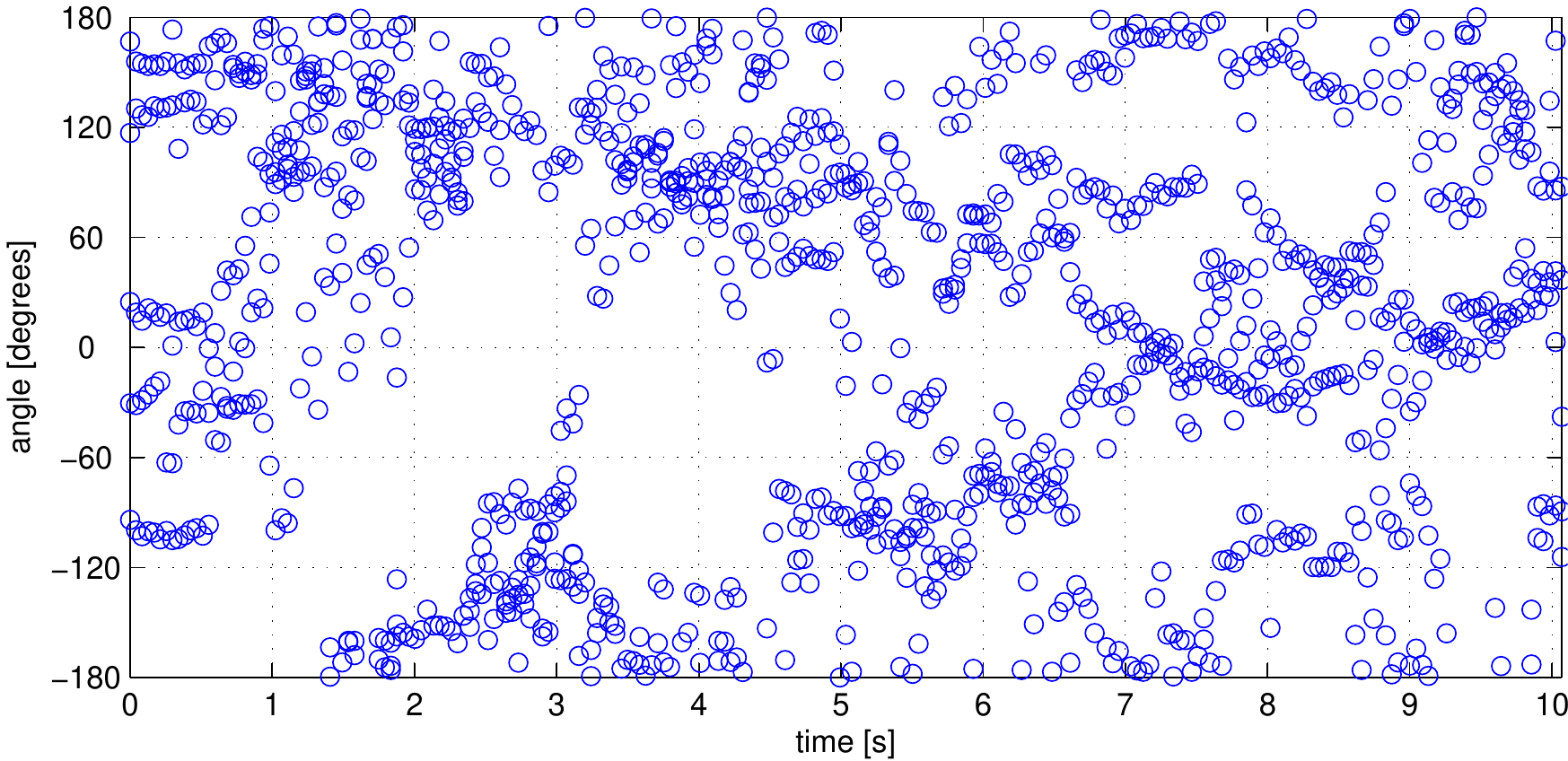}
	%\caption{}
  %\label{fig:WGMMmeans}
	\end{subfigure}
  \centering
	\begin{subfigure}[b]{0.5\linewidth}
  \includegraphics[width = 1.0\linewidth]{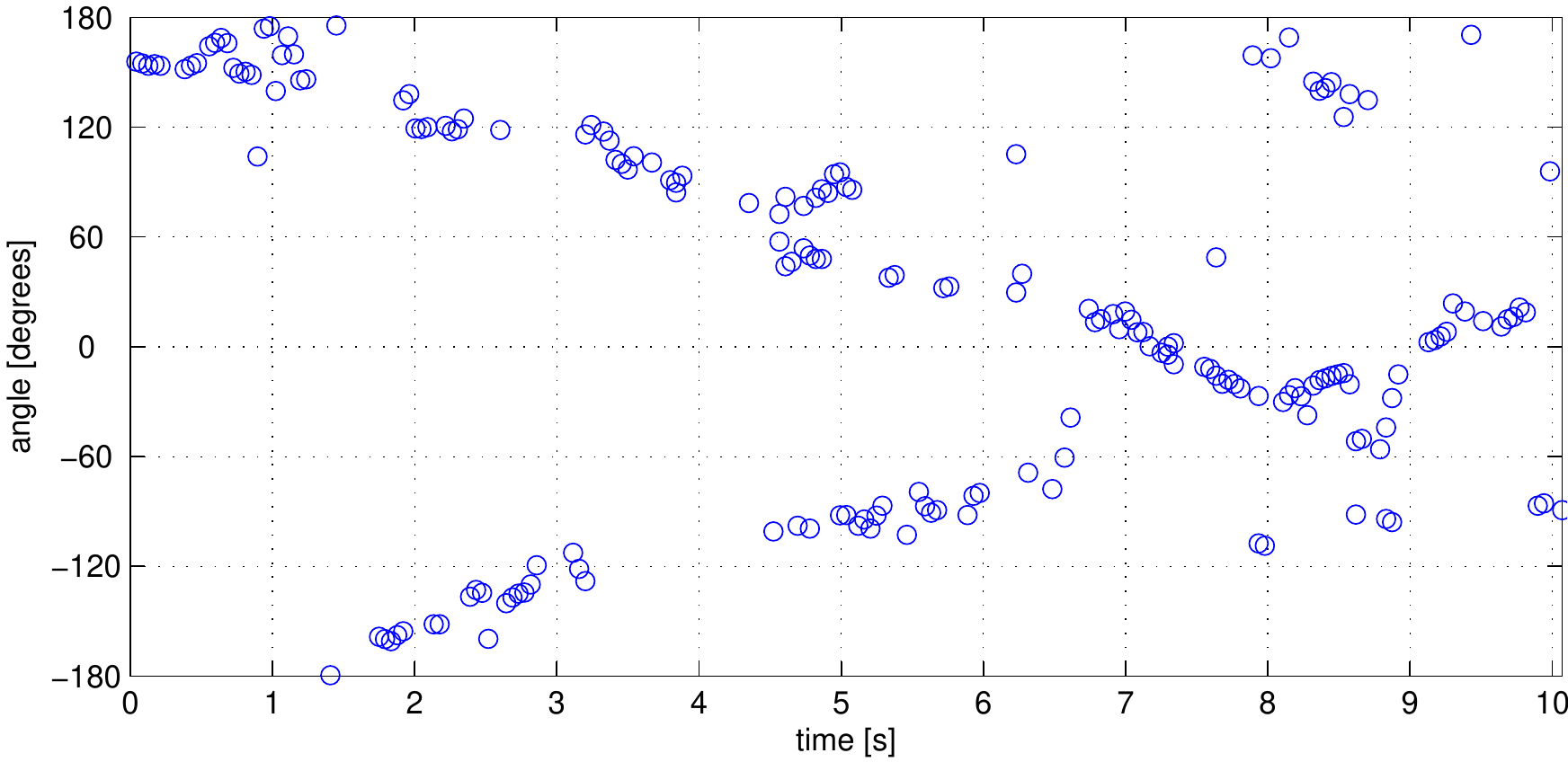}
	%\caption{ }
  %\label{fig:WGMMmeans2}
	\end{subfigure}
	\caption{Upper panel illustrates all estimated WGMM means $\Zmu$ for each time-frame $\jind$. In the lower panel WGMM means after removing measurements with $\Zsigma > 0.6 \; \mathrm{rad} \; (\approx 35^\circ)$ or $\Za < 0.15$ are illustrated}
  \label{fig:WGMMmeans}
\vspace{-4mm}
\end{figure}

%With the above definitions for state and measurement vectors, Equations \eqref{eq:measvec} and \eqref{eq:statevec}, the measurement model and the state transition model become linear. 

\subsubsection{Multi-target DOA tracking implementation}
For the actual RBPF implementation we now reintroduce the WGMM component index $\Gind$ and the source index $\qind$. % and the association of $\Gind$th to one of the detected source $\qind = 1, \dots, \qmax$ is estimated withing the RBPF algorith \cite{sarkka1}. 
The state vector is defined individually for each detected source $\qind = 1,\dots,\qmax$ and is denoted hereafter as $\state_\jind^{(\qind)}$. Similarly the multiple measurements at same time step obtained from the WGMM model are denoted by $\measur_\jind^{(\Gind)} = \begin{bmatrix} \cos(\Zmu), \sin(\Zmu) \end{bmatrix}^T$

The RBPF implementation \cite{hartikainen} is applied to the measurements $\measur_\jind^{(\Gind)}$ with measurement noise $\measnoise_{\jind}^{(\Gind)} \sim \mathcal{N}(0, \Zsigma \mathbf{I})$. The multi-target tracker detects the sources and makes the association of $\Gind$th WGMM measurement belonging to one of sources $\qind = 1,\dots,\qmax$. Alternatively, if none of the active source particle distributions indicate a probability higher than the clutter prior probability, then the current measurement is regarded as clutter. The clutter prior probability is a fixed pre-set value to validate the minimum threshold when the observed measurement is linked to existing source.

%Alternatively, if none of the active source statistics indicate a probability higher than clutter prior probability, then the measurement is labeled as clutter. %The source indices ($\qind = 1,\dots,\qmax$) after the particle filter and in the CNMF model in Equation \eqref{eq:CNMFmodel2} are same due to notational conformity. %It should be noted that the number of underlying sources present in the mixture is unknown at the first CNMF stage, and thus a small constant is used to estimate initial spatial weights which are used to produce mixture spatial weights $\zoj$.

The output of the tracker is the state of each source $\qind$ at each time frame, denoted by $\state_\jind^{(\qind)}$. %It should be noted that not all sources are active at each time frame. 
Extracting the DOA from the tracked source state requires calculating the angle of the vector defined by the 2-D coordinates and thus the resulting DOA trajectories are obtained as
\begin{equation}
\label{eq:PFtoAngle}
\Zmup \leftarrow \atan2(\state_{\jind,2}^{(\qind)} / \state_{\jind,1}^{(\qind)}).
\end{equation}

\begin{figure}[tp]
  \centering
  \includegraphics[width = 0.5\linewidth]{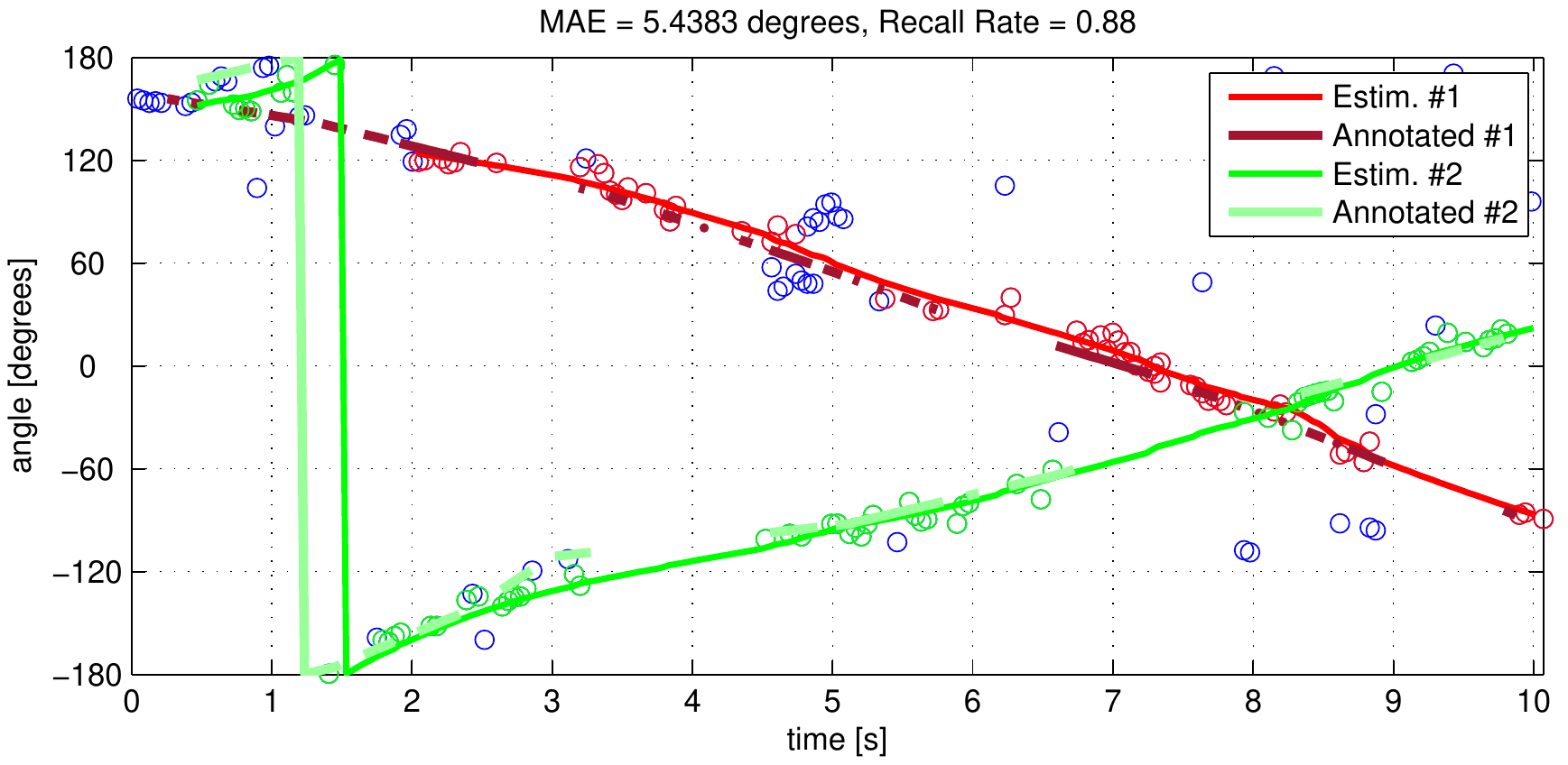}
	\caption{The acoustic tracking result of two sources intersecting and ground truth annotations illustrated when the source is active (voice activity detection by energy thresholding using signal from close-field microphone).}
  \label{fig:PFresult}
\vspace{-4mm}
\end{figure}

The tracking result for a one test signal is illustrated in Figure \ref{fig:PFresult}, where the input of the acoustic tracking are the ones depicted in the bottom panel of Figure \ref{fig:WGMMmeans}. The test signal is chosen such that it shows two problematic cases, sources start from the same position and intersect at 8 seconds going to the opposite directions. The tracking result indicates that the second source is detected at 2 seconds from the start just when the sources have traveled far enough from each other, resulting to approximately one first word being missed from the second speaker. The tracker is able to maintain the source association and track correctly the trajectories of intersecting sources. 
%The particle filter has correctly detected two sources and the tracked state converted back to azimuth angle by Equation \eqref{eq:PFtoAngle}. However, the two sources starting from approximately same position has caused the decetion of second source not until its second spoken word around 4 seconds mark. 

\section{Separation model}
\label{sec:CNMF}

\subsection{Mixture in spatial covariance domain}
\label{sec:Covdomain}
%In audio signal processing with microphone arrays the channel-wise phase and level differences encode the information regarding the spatial location of the sound sources. 
For the separation part we represent the microphone array signal using mixture SCMs $\Xcov \in \mathbb{C}^{\mmax \times \mmax}$. We use magnitude square rooted version of the mixture STFT obtained as, 
\begin{equation}
\label{eq:xmagsquaredroot}
\xFDhat = [|x_{\iind\jind 1}|^{1/2} \, \mathrm{sign}(x_{\iind\jind 1}), \dots , |x_{\iind\jind\mmax}|^{1/2} \, \mathrm{sign}(x_{\iind\jind\mmax})]^T
\end{equation}
where $\mathrm{sign}(z) = z/|z|$ is the signum function for complex numbers. %, i.e. $|\xFDhat| = \sqrt{|\xFD|}$ (see, for example, \cite{sawadaCNMFwaspaa,nikunenCNMFtaslp}). 

The mixture SCM is calculated as $\Xcov = \xFDhat \xFDhat^H$ for each TF-point ($\iind,\jind$). The diagonals of each $\Xcov$ contains the magnitude spectrogram of each input channel. The argument and absolute value of $[\Xcov]_{\mind_1,\mind_2}$ (off-diagonal values) represents the phase difference and magnitude correlation, respectively, between microphones $(\mind_1,\mind_2)$ for a TF-point ($\iind,\jind$). 

The TF domain mixing in Equation \eqref{eq:FDmodel} can be approximated using mixture SCMs as
\begin{equation}
\label{eq:covmodel}
\Xcov \approx \Xcovhat = \sum_{\qind=1}^\qmax \Hmat \sFDhat,
\end{equation}
where %$\sFDhat = (\sFD\overline{\sFD})^{(1/2)}$ 
$\sFDhat = \sqrt{ \sFD\sFD^*}$ is positive real-valued magnitude spectrogram of source $\qind$ and $\Hmat = \hFD \hFD^H / ||\hFD \hFD^H ||_F$ are the SCMs of the frequency domain RIRs $\hFD$. The mixing Equation \eqref{eq:covmodel} is hereafter referred to as spatial covariance domain mixing.

%uses weighted sum of entities called "DOA kernels" each containing a phase difference caused by a single direction vector 

%is composed of a weighted sum of entities called "DOA kernels" each containing a phase difference caused by a single direction vector.
%The DOA based SCM model has been shown to exceed the performance of the direct estimation in \cite{nikunenCNMFtaslp,nikunenCNMFicassp} given that the same optimization criterion is used for both alternatives. 

%In the proposed algorithm first a preliminary estimate of the source mixture magnitude spectrum and mixture directions weights is estimated, referred to as first CNMF stage. The tracking is applied to solve data association and discrimination of different acoustical source based on the mixture direction weights. The result of the tracking is used for initializing the direction weights of each source at each time frame based on the tracked result and re-estimating the optimal magnitude model parameters.

%\subsection{CNMF Model for Time-Variant Mixing}
%Usually, the duration of the audio signal segment processed using a CNMF is of order of several seconds. All previous CNMF formulations have effectively estimated the SCMs for each separated source as an average over the analyzed segment. 
\subsection{Multichannel NMF model with time-variant mixing}
\label{sec:CNMFmodel}
The proposed algorithm uses multichannel NMF for source spectrogram estimation and it is based on alternating estimation of the source magnitude spectrogram $\sFDhat$ and its associated spatial properties in the form of the SCMs $\Hmat$. In all previous works \cite{ozerovNTFmulti,sawadaCNMFtaslp,nikunenCNMFicassp} the problem definition has been simplified for stationary sound sources and the SCMs being fixed for all STFT frames $\jind$ within the analyzed audio segment. Here we present a novel extension of the multichannel NMF model for time-variant mixing.

In multichannel NMF the model for magnitude spectrogram is equivalent to conventional NMF, which is composed of fixed spectral basis and their time-dependent activations. The SCMs can be %estimated 
unconstrained \cite{sawadaCNMFwaspaa} or as proposed in the earlier works of the authors \cite{nikunenCNMFtaslp,nikunenCNMFicassp} based on a model that represents SCMs as a weighted sum of entities called "DOA kernels" each containing a phase difference caused by a single direction vector. This ensures SCMs to comply with the array geometry and match with the time-delays the chosen microphone placement allows. %causes.

The NMF magnitude model for source magnitude spectrogram is given as 
\begin{equation}
\label{eq:NMFmodel}
\sFDhat \approx \sum_{\kind=1}^{\kmax} \bkp \tik \vkj, \quad \bkp, \tik ,\vkj \geq 0. 
\end{equation}
Parameters $\tik$ over all frequency indices $\iind = 1,\dots,\imax$ represent the magnitude spectrum of single NMF component $\kind$, and $\vkj$ denotes the component gain in each frame $\jind$. One NMF component represents a single spectrally repetitive event estimated from the mixture and one source is modeled as a sum of multiple components. Parameter $\bkp \in [0,1] $ represents a weight associating NMF component $\kind$ to source $\qind$. The soft valued $\bkp$ is motivated by different types of sound sources requiring different amount of spectral templates for accurate modeling and learning of the optimal division of components is determined through parameter updates. For example, stationary noise can be represented using only a few NMF components whereas spectrum of speech varies over time and requires many spectral templates to be modeled precisely. Similar strategy for $\bkp$ is used for example in \cite{sawadaCNMFtaslp}. Typically the final values of $\bkp$ are mostly binary and only few number of NMF components are shared among sources.

The multichannel NMF model with time-variant mixing can be trivially derived from the SCM mixing Equation \eqref{eq:covmodel} by substituting the above defined NMF model \eqref{eq:NMFmodel} in it, resulting in
\begin{equation}
\label{eq:CNMFmodel}
\Xcov \approx \Xcovhat = \sum_{\qind = 1}^{\qmax} \Hmat \underbrace{\big[\sum_{\kind=1}^{\kmax} \bkp \tik \vkj \big]}_{\approx \sFDhat}.
\end{equation}
%The multiplicative updates for finding the optimal parameters of the CNMF model \eqref{eq:CNMFmodel} can be derived as proposed in \cite{sawadaCNMFtaslp} and the optimization criterion of squared Frobenius norm or Itakura-Saito divergence can be used.

%It is evident that, regardless of the choice of the cost function, optimizing $\Hmat$ individually for each time frame $\jind$ will not constrain the spatial behavior of the estimated sources over time. The spatial focus of source $\qind$ in the model may change from frame to frame between the actual sound sources.  Additionally, the spatial evidence in one frame is noisy, which would lead to estimates of SCMs in individual frames also noisy and unreliable.
 
\subsection{Direction of arrival -based SCM model}
\label{sec:SCMmodel}
In order to constrain the spatial behavior of sources over time using information from the acoustic tracking approach or other prior information, the SCMs need to be interpreted by spatial location of each source in each time frame. The SCM model proposed in \cite{nikunenCNMFtaslp} parametrizes stationary SCMs based on DOA and here we extend the model for time-variant SCMs $\Hmat$. % The DOA based SCM model allows interpretation of source DOA in each time frame. Additionally, the proposed model also estimates the SCMs independent of frequency which alleviates phase ambiguity caused by the spatial aliasing.

%\subsubsection{Direct-path propagation}
%We define a look direction vector $\mathbf{k}_\oind$ pointing towards direction parametrized by azimuth $\theta_\oind \in [0, 2\pi]$ and elevation $\varphi_\oind \in [0, \pi]$ originating from the geometric center of the array. In a general case the look direction vectors indexed by $\oind$ would sample the space around the array approximately uniformly, but in this paper we only consider the zero elevation plane, \ie, $\varphi_\oind = 0$. We assume that the sources of interest lie approximately on the xy-plane with respect to the microphone array and the directional statistics used in tracking of the sources simplifies to a univariate case when the sampling of the spatial space around the array is by azimuthal information only. A sparse grid of look directions vectors with spacing of adjacent azimuths by $pi/12$ are illustrated in Figure \ref{fig:leikkaus} along with an array casing and microphone positions corresponding to the actual compact array used in the evaluations. %Additionally, the wrapped Gaussian mixture model can be defined only for 1-dimensional data. 

Converting the TDOA in Equation \eqref{eq:TDoA} to a phase difference results in DOA kernels $\Wmat \in \mathbb{C}^{\mmax \times \mmax}$ for a microphone pair $(m_1,m_2)$ defined as
\begin{equation}
\label{eq:Wkernel}
[\mathbf{W}_{\iind\oind}]_{m_1,m_2}= \exp \big( j \omega_\iind \tau_{\oind}({m_1,m_2}) \big), % \quad f_\iind = (\iind-1) F_s/{N},
\end{equation}
where %$\fHz_\iind$ is the frequency of $\iind$th DFT bin index and 
$\tau_\oind({m_1,m_2})$ denotes the time delay caused by a source at a direction $\mathbf{k}_\oind$. 

%\subsubsection{SCM model of Time-Variant Mixing}
A linear combination of DOA kernal gives the model for time-varying source SCMs defined as
\begin{equation}
\label{eq:SCMmodel}
\Hmat = \sum_{\oind=1}^\omax \Wmat \zpoj.
\end{equation}
The directions weights $\zpoj$ denote the spatial location and spread of the source $\qind$ at each time frame $\jind$. %However, the remaining problem in free optimization ($ \min \sum_{\iind=1}^{\imax} \sum_{\jind=1}^{\jmax} || \Xcov- \Xcovhat||^2_F$)  is that the spatial behavior of the sources is not constrained over time and the actual acoustical sources can be permutated from frame-to-frame. This problem is addressed later in the paper in the form of acoustical source tracking and solving the data association in Section \ref{sec:tracking}. 

The direction weights $\zpoj$ can be interpreted as probabilities of source $\qind$ originating from each direction $\oind$. In anechoic conditions only one of the directions weights $\zpoj$ in each time frame would be nonzero, \ie, the direct path explains all spatial information of the source, however in reverberant conditions several of the direction weights are active.

%\begin{figure}[tb]
  %\centering
  %\includegraphics[width = 0.7\linewidth]{pallo2}
 %\caption{The look directions are illustrated as red circles and the surface defined by the direction weights being all one (	$\zpoj = 1$).}
 %\label{fig:pallo}
%\end{figure}

%\subsection{Initialization of spatial weights and background noise modeling}
%\label{sec:restore}

%The use of CNMF model in Equation \eqref{eq:CNMFmodel} for source separation requires defining $\Hmat$ via the DOA kernels $\Wmat$ and spatial weights $\zpoj$ denoting the source movement over time. 
%The spatial weights $\zpoj$ can be initialized according to estimated source movement over time. 

%The DOA based SCM model \eqref{eq:SCMmodel} allows interpreting the source location in terms of discrete DOA distribution at each given time frame. This enables to incorporate source tracking information along the CNMF parameter estimation framework by setting the spatial weights $\zpoj$ in such way that each source in the CNMF model corresponds to spatial evidence of each tracked source.

%The individual steps for setting the spatial weights for the CNMF parameter estimation is illustrated in Figure \ref{fig:Zfigures} including the mixture spatial weights (SRP) and the WGMM model of them. The procedure for initializing the spatial weights is explained in following paragraphs. 

The spatial weights corresponding to tracked sources and their DOA trajectories are set using the wrapped Gaussian distribution as
\begin{equation}
\label{eq:Zrestore}
\zpojt = \mathcal{N}_w  (\theta_\oind ; \Zmup, \Zsigmap),
\end{equation}
where $\Zmup$ is obtained as specified in Equation \eqref{eq:PFtoAngle} and the variance $\Zsigmap$ is used to control the width %of the restored Gaussian and thus the width 
of the spatial window of source $\qind$ in the separation model. The goal is to have a large spatial window when the tracker is certain and the output suppressed (small spatial window) when no new measurements from the predicted source position are observed and the tracker state indicates high uncertainty. This strategy is motivated by ensuring that small untracked deviations in source movement do not cause the spatial focus of the separation to veer off momentarily from the target source and lead to suppression of the desired source content. In experimental tests the source spectrogram estimation by multichannel NMF proved to be sensitive to small errors if the spatial window used was very small. The small spatial window in case of no source activity can be motivated from similar perspective, the estimated source spectrogram from very constrained spatial window is less likely to capture spectrogram details of other sources close to the target trajectory.

The acoustic tracker output variance denoted as $\sigma_{\jind,\qind}^2$ at each time step $\jind$ indicates the uncertainty of the source being present at its respective predicted direction $\Zmup$. The above specified strategy can be obtained from the tracker output variance by specifying $ \Zsigmap = c - \sigma_{\jind,\qind}^2 $ with a constant $c = \max\limits_{\jind,\qind}(\sigma_{\jind,\qind}^2) + \min\limits_{\jind,\qind}(\sigma_{\jind,\qind}^2)$. The operation maps the maximum output variance to the smallest spatial window and vice versa. In practice value range of $\sigma_{\jind,\qind}^2$ is restricted to avoid specifying extremely wide and narrow spatial windows and thus the value of constant $c$ is set in advance. The limits for $\sigma_{\jind,\qind}^2$ are discussed in more details in Section \ref{sec:procparam}.
%Each source trajectory at each time frame  consist of a single wrapped Gaussian and thus the weights $a_\jind$ are effectively equal to one and are omitted from Equation \eqref{eq:Zrestore}.

%The tracker output variance $\sigma_{\jind,\qind}^2$ at each time step $\jind$ indicates the uncertainty of the source being present at their respective predicted direction and the variance is used to control the width of the Gaussian and thus the width of the spatial window of the separation. 
%Using the variance of $-\sigma_{\jind,\qind}^2 + c$ as specified above with a constant $c = \max(\sigma_{\jind,\qind}^2) + \min(\sigma_{\jind,\qind}^2)$ equals 

\begin{figure}[tb]
  \centering
  \includegraphics[width = 0.5\linewidth]{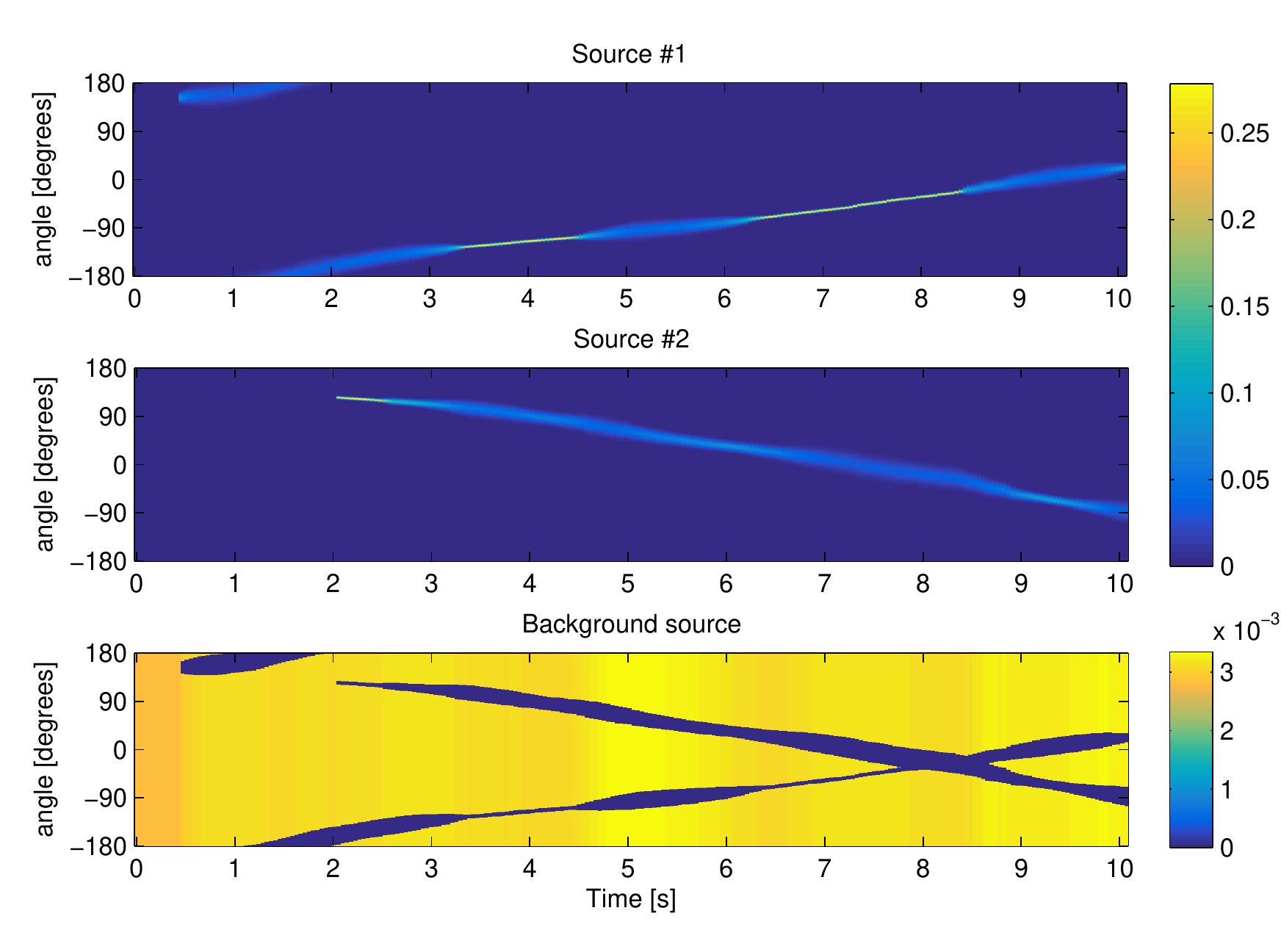}
	\caption{The reconstructed spatial weights as given in Equation \eqref{eq:Zrestore} for two detected sources are illustrated in two top panels and the spatial weights corresponding to the background source are illustrated in the bottom panel.}
  \label{fig:Background}
\vspace{-4mm}
\end{figure}

The direction weights for each source at each time frame are scaled to unity $l_1$-norm ($\sum_{\oind=1}^\omax \zpojt = 1$). This is done to restrict $\Hmat$ to only model spatial behavior of sources and not affect modeling the overall energy of the sources. Additionally, when the source is considered inactive, \ie,  before its birth or after its death, all the direction weights in the corresponding time frame are set to zero. The spatial weights $\zpojt$ corresponding to the tracking result in Figure \ref{fig:PFresult} are illustrated in two top panels of Figure \ref{fig:Background}. By comparing to Figure \ref{fig:PFresult}, it can be seen that when no new measurements are observed the tracker output state variance is high and the spatial weights are concentrated tightly around the mean, whereas in case of high certainty the spatial spread is wider.

%The bottom panel in Figure \ref{fig:Zfigures} illustrates the situation after equation \eqref{eq:Zrestore} using $\sigma^2 = 0.2$.

 % Otherwise, \ie, the source is inactive, all the direction weights are set to zero. During the active segments the equivalence of $\sum_{\qind} \sFDhat = \sFDtilde$ holds.
%The variance could be retrieved from the particle filter statistics (the variance of the x and y position at each time step), but it is not considered in this work for the sake of simplicity.

In order to model the background noise and diffuse sources, an additional background source is added with direction weights set to one in indices where $\sum_\qind \zpojt < \zth$ and zero otherwise. The threshold $\zth$ is set to allow the detected and tracked sources to capture all spatial evidence within approximately +-30 degrees from their estimated DOAs when the certainty for the given source is high. With the chosen background modeling strategy the tracked sources have exclusive prior to model signals originating from the tracked DOA, with the exception of two DOA trajectories intersecting. %and thus both are active at the same direction indices. 
An example of the background spatial weights is illustrated in bottom panel of Figure \ref{fig:Background}. Note that the differently colored regions at different times are due to the scaling ($\sum_{\oind=1}^\omax \zpojt = 1$) and different spatial window widths of the tracked sources at the corresponding time indices.

%\begin{figure}[tb]
  %\centering
  %\includegraphics[width = 0.95\linewidth]{journal_SRP_and_WGMM}
	%\caption{Upper panel illustrates the SRP and lower panel illustrates the WGMM model estimated from it.}
  %\label{fig:Zfigures}
%\end{figure}

\subsection{Parameter Estimation}
\label{sec:CNMFparamest}

The multichannel NMF model \eqref{eq:CNMFmodel} with the time-varying DOA kernel based SCM model \eqref{eq:SCMmodel} and the spatial weights as specified in Equation \eqref{eq:Zrestore} result to model
\begin{equation}
\label{eq:CNMFmodel2}
\Xcov \approx \Xcovhat = \sum_{\qind = 1}^{\qmax} \underbrace{\sum_{\oind=1}^\omax \Wmat \zpoj}_{\Hmat} \underbrace{\sum_{\kind=1}^{\kmax} \bkp \tik \vkj }_{\approx \sFDhat}.
\end{equation}
In order to use the above model for separation, parameters $\bkp, \tik$ and $\vkj$ defining the magnitude spectrogram modeling part need to be estimated with respect to appropriate optimization criterion.
%After the spatial weights for the SCM model \eqref{eq:SCMmodel} are set the updates of the CNMF model specified in Section \ref{sec:CNMFparamest} are repeated for fixed amount of iterations. 
%This corresponds to estimation of the source spectrogram model for the found DOA trajectories and spectral evidence originating from the direction specified by the acoustic tracking. %The spatial weights after the second CNMF stage are illustrated in the lowermost panel of Figure \ref{fig:Zfigures}. 

%The non-negative parameters to be optimized are $\bkp, \tik$ and $\vkj$. It is assumed at this point that $\Hmat$ is set extrenally (defined using equation \eqref{eq:Wkernel} and $\zpoj$ is obtained by acoustical tracking) and it is kept fixed during the estimation of ${\sFDhat}$. %Additionally the absolute value of $\Wmat$ needs to be estimated while retaining original argument of each of its element. %The magnitudes of DOA kernels $\Wmat$ model the magnitude differences between each channel and sources can have different gain with respect to each microphone depending on the source DOA.

We use the squared Frobenius norm as the cost function, defined as $\sum_{\iind=1}^{\imax} \sum_{\jind=1}^{\jmax} || \Xcov- \Xcovhat||^2_F$. Multiplicative updates for estimating the optimal parameters in an iterative manner can be obtained by partial derivation %of the total modeling criterion, \ie, 
of the cost function and use of auxiliary variables as in expectation maximization algorithm \cite{EMalgo}. 
The procedure for obtaining multiplicative updates for different multichannel NMF models and optimization criteria are proposed and presented in \cite{sawadaCNMFtaslp} and can be extended for the new proposed formulation in \eqref{eq:CNMFmodel2}. %utilized for example in \cite{nikunenCNMFtaslp,nikunenCNMFicassp}. 
The entire probabilistic formulation is not repeated here and can be reviewed from~\cite{sawadaCNMFtaslp}.

The update equations for the non-negative parameters are
\begin{align}
\label{eq:updatesmulb}
\bkp & \leftarrow \bkp \frac{\sum_{\iind,\jind} \tik \vkj \mathrm{tr}(\Xcov \Hmat)}{\sum_{\iind,\jind} \tik \vkj  \mathrm{tr}(\Xcovhat \Hmat)},
\end{align}
\begin{align}
\label{eq:updatesmult}
\tik & \leftarrow \tik \frac{\sum_{\jind,\qind} \bkp \vkj \mathrm{tr}(\Xcov \Hmat)} {\sum_{\jind,\qind} \bkp \vkj \mathrm{tr}(\Xcovhat \Hmat)}, \\
\label{eq:updatesmulv}
\vkj & \leftarrow \vkj \frac{\sum_{\iind,\qind} \bkp \tik \mathrm{tr}(\Xcov \Hmat)}{\sum_{\iind,\qind} \bkp \tik \mathrm{tr}(\Xcovhat \Hmat)}.
\end{align}
Note that in contrast to earlier works on multichannel NMF for separation of stationary sound sources \cite{sawadaCNMFtaslp,nikunenCNMFtaslp}, we do not update the SCM part $\Hmat$. It is assumed that the acoustic source tracking and spatial weights of the DOA kernels $\Wmat$ fully represent the spatial behavior of the source. This strategy is assessed in more details in discussion Section \ref{sec:discussion}. %After estimation of the source magnitude spectrogram model $\sFDhat = \sum_{\kind=1}^{\kmax} \bkp \tik \vkj $ the separation is carried out by applying Wiener filtering in Equation \eqref{eq:recon2}.

\subsection{Source separation}
For extracting the source signal from the mixture we use combination of single-channel Wiener filter and delay-and-sum beamforming. Separation soft mask ${m}_{\iind\jind,\qind}$ for extracting the source spectrogram from the mixture are obtained using the estimated real valued magnitude spectrogram $\sFDhat$ to formulate a generalized Wiener filter defined as
%Source separation from the CNMF model is achieved by defining separation masks ${m}_{\iind\jind,\qind}$ based on the estimated real valued magnitude spectrogram $\sFDhat$   % and estimation of otherwise defining SCMs $\Hmat$ to denote the source movement over time. These estimated quantities are used reconstructing the source signals by multichannel Wiener filtering
%\begin{equation}
%\label{eq:recon1}
%\yFDhat = \sFDhat \Xcovhat^{-1} \Hmat \xFD,
%\end{equation}
%by means of Wiener filtering,
\begin{equation}
\label{eq:recon2}
\yFDhat = m_{\iind\jind,\qind} \xFD = \frac { \sFDhat } { \sum_{\qind} \sFDhat } \xFD .
\end{equation}
%The differences of the reconstruction methods are discussed in \cite{sawadaCNMFtaslp} with no clear conclusion on favoring another.
We employ delay-and-sum beamforming to produce single channel source signal from the separated multichannel signals $\yFDhat$ (having the mixture signal phase). The final estimate of the sources are given as
\begin{equation}
\label{eq:DSB}
\yFDsingle = \mathbf{w}_{\iind\jind,\qind}^H \yFDhat, 
\end{equation}
where $\mathbf{w}_{\iind\jind,\qind} $ are the DSB weights (steering vector%$[0, \exp (j \omega_\iind \tau_\oind(1,2)), \dots, \exp (j \omega_\iind \tau_\oind(1,\mmax))]^T$ 
) towards estimated direction of source $\qind$ at time frame $\jind$. %The steering vector is discussed later in the paper in more details.
%$\mathbf{w}_{\iind\qind}^H = [ e^{ - j \omega_\iind \tau_{11} (\mathbf{\hat{k}}_\qind)}, \dots e^{ - j \omega_\iind \tau_{1m} (\mathbf{\hat{k}}_\qind)}]$,
Finally, the time-domain signal are reconstructed by applying inverse DFT to each frame which are further combined using overlap-add processing.

%\includegraphics[width = 0.8\linewidth]{lohko}
%\psfrag{spa1}{$\zpo$}
%\psfragfig[width=0.8\linewidth]{lohko}

\begin{comment}
\begin{table}[tb]
\caption{Microphone locations.}
\label{tab:micpos}
\centering
\begin{tabular}{|c|ccc|}
  \hline
  \textbf{Mic} & \textbf{x (mm)} & \textbf{y (mm)} & \textbf{z (mm)} \\
  \hline
  \textbf{1} & 0 & -46 & 6 \\
  \textbf{2} & -22 & -8 & 6 \\
  \textbf{3} & 22 & -8 & 6 \\
  \textbf{4} & 0 & 61 & -18 \\
  \hline
\end{tabular}
\vspace{-4mm}
\end{table}
\end{comment}

\section{Evaluation}
\label{sec:Eval}
In this section we present the objective separation performance and tracking performance of the proposed method using real recordings of moving sound sources. Additionally, we evaluate the separation performance of the proposed algorithm in a setting where the sources are not moving allowing comparison to conventional spatial and spectrogram factorization models assuming stationary sources. % \cite{sawadaCNMFtaslp,nikunenCNMFtaslp,nikunenCNMFicassp}.

\begin{figure}[tb]
  \centering
  \includegraphics[width = 0.40\linewidth]{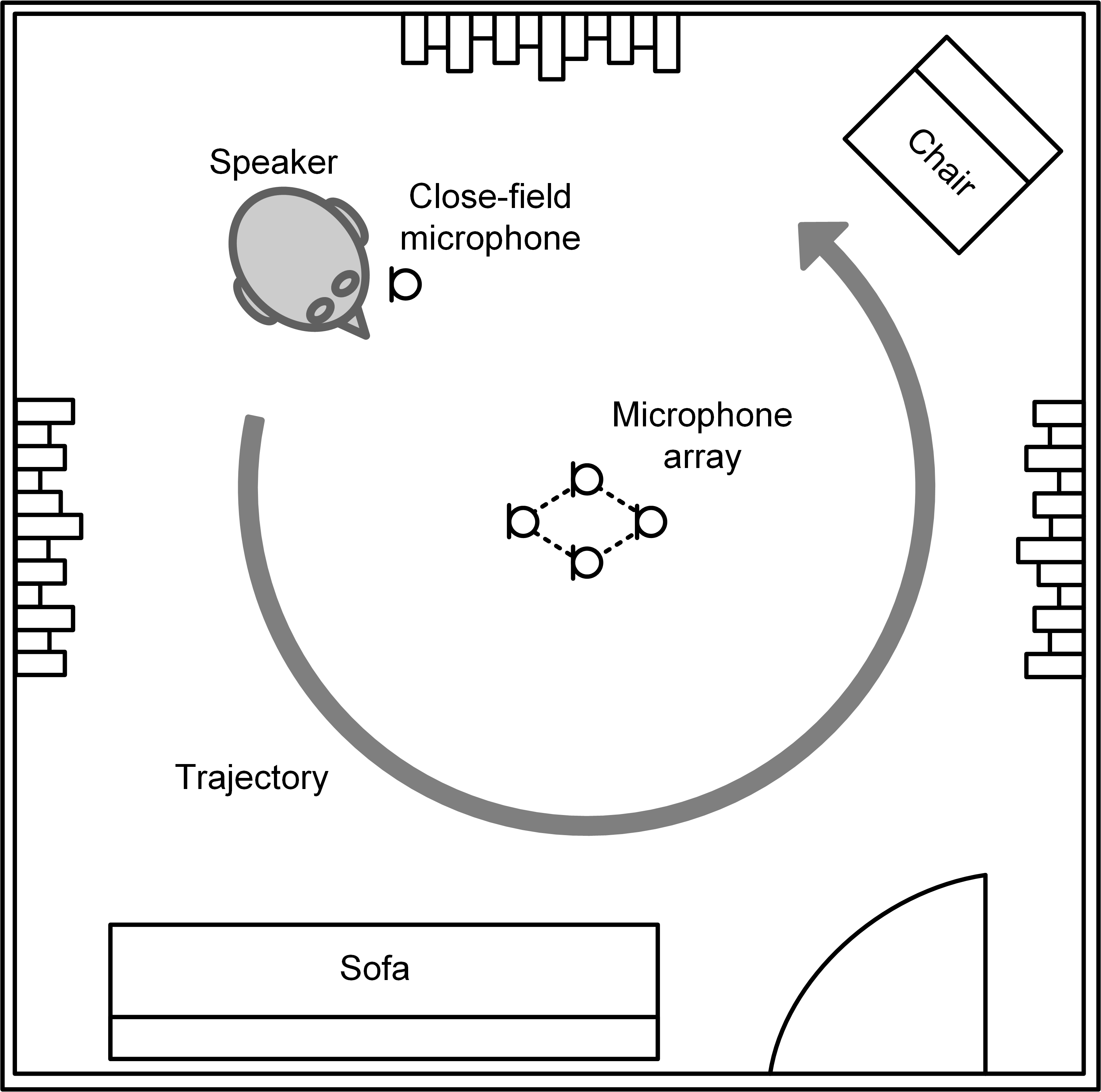}
	\caption{Illustration of the recording setup and the source movement.}
  \label{fig:recfig}
\vspace{-3mm}
\end{figure}

%\subsection{Datasets for Development and Evaluation}

\subsection{Datasets with moving sources}
\label{sec:datasets}
The development and evaluation material was recorded using a compact microphone array consisting of four Sennheiser MKE2 omnidirectional condenser microphones placed on a diamond pattern illustrated in \ref{fig:leikkaus} and exact locations of microphones are documented in \cite{nikunenCNMFtaslp}. The recordings were conducted in an acoustically treated room with dimensions $4.53$ m $\times$ $3.96$ m $\times$ $2.59$ m and reverberation time $T_{60} =$ 0.26 s. The microphone array was placed approximately at the center of the room. Four persons spoke phonetically balanced sentences while walking clockwise (CW) and counterclockwise (CCW) around the microphone array at approximately constant velocity and at average distance of one meter from the array center. The overall assembly of the recordings and the movement of the source is illustrated in Figure \ref{fig:recfig}.

Two CW and two CCW 30-second recordings with four persons were done totaling to 16 signals. %The persons never changed their direction within one recording. 
All speakers started from the same position and walked on average two times around the array within the recorded 30-second segment. Recordings were done individually allowing producing mixture signals by combining the recordings from different persons. Reference speech signal was captured by a close-field microphone (AKG C520). Additionally, 16 recordings with a loudspeaker playing babble noise and music outside the recording room with the door open were done and considered as a stationary (S) sound source with highly reflected propagation path. The movement of each individual speaker was annotated by hand based on SRP. %  and using a video of the recording session as a reference. 
An example of annotations is illustrated in Figure \ref{fig:PFresult}. Note that the annotations are only plotted when the source is active (a simple energy threshold from the close-field microphone signal). The VAD information is only used for evaluation purposes and not by the proposed algorithm. %even though they are continuous and used as such for the evaluation of tracking performance.

Three different datasets were generated, one for development and two for evaluation purposes by mixing two and three individual speaker recordings. All mixture utterances in all datasets were 10 seconds in duration. In all datasets the signals were manually cut in such way that speaker trajectories based on the annotations were no closer than $45^\circ$ when going in the same direction (CW vs. CW and CCW vs. CCW). Naturally, the trajectories can intersect in the case of opposite directions (CW vs. CCW). %The SNR between speakers was not equalized, since all the recordings had approximately same overall loudness.

For the development set the first 15 seconds from each recording were used, while the remaining 15 to 30 seconds were used to generate the evaluation sets. In the development set 8 mixtures of two speakers and 16 live recordings were generated and each recording was only used once. The first evaluation dataset consists of 48 mixtures of two speakers using all possible unique combinations of the recordings with different speakers. The second evaluation dataset contains 16 mixtures of three speakers based on a subset of all possible unique combinations. The subset was chosen to represent all different source trajectory combinations (all sources moving in the same direction vs one of the sources moving in the opposite direction). The datasets are summarized in Table \ref{tab:dataset}. 
%with description given in table at page 3 in project meeting slide \url{nokia050614.pdf}. % Add the table to this document 

The global parameters related to tracking and separation performance of the proposed algorithm were optimized using the development dataset. The recorded signals were downsampled and processed with sampling rate of $F_s = 24000$ Hz. 

%& \multicolumn{2}{c}{Movement Type} & \\ 
\begin{table}
\setlength{\tabcolsep}{4pt}
\caption{Description of datasets.}
\label{tab:dataset}
\begin{tabular}{cccc}
\multicolumn{4}{l}{\textbf{Developmenet dataset}} \\
\toprule
Number of & & & \\
samples & Source 1 & Source 2 & Details \\ 
\midrule
2 & CW & CW & $\measuredangle > 45^\circ$ \\
2 & CCW & CCW & $\measuredangle > 45^\circ$ \\
4 & CW & CCW & Sources intersect \\ 
12 & CW/CCW & S (babble) & SNR = -5, -10 and -15 dB \\
4 & CW/CCW & S (music) & SNR = -10 dB \\
\bottomrule
& & & \\
\end{tabular}

\begin{tabular}{cccc}
\multicolumn{4}{l}{\textbf{Evaluation dataset, 2 sources}} \\
\toprule
Number of & & & \\
samples & Source 1 & Source 2 & Details \\ 
\midrule
16 & CW & CW & $\measuredangle > 45^\circ$ \\
16 & CCW & CCW & $\measuredangle > 45^\circ$ \\
16 & CW & CCW & Sources intersect \\
\bottomrule
& & & \\
\end{tabular}

\begin{tabular}{ccccc}
\multicolumn{5}{l}{\textbf{Evaluation dataset, 3 sources}} \\
\toprule
Number of & & & & \\
samples & Source 1 & Source 2 & Source 3 & Details \\ 
\midrule
4 & CW & CW & CW & $\measuredangle > 45^\circ$ \\
4 & CCW & CCW & CCW  & $\measuredangle > 45^\circ$ \\
4 & CW & CW & CCW &  Sources intersect \\
4 & CCW & CCW & CW & Sources intersect \\
\bottomrule
& & & \\
\end{tabular}

\begin{tabular}{cccc}
\multicolumn{4}{l}{\textbf{Dataset with stationary sources from \cite{nikunenCNMFtaslp}}} \\
\toprule
Number of & & &\\
samples & Source 1 & Source 2 & Details \\ 
\midrule
8 / 8 & $45^\circ$ / $135^\circ$ & $90^\circ$ / $180^\circ$ & $\measuredangle = 45^\circ$ \\
%8/8 & /$135^\circ$ & $180^\circ$ \\
8 / 8 & $0^\circ$ / $45^\circ$ & $90^\circ$ / $135^\circ$ & $\measuredangle = 90^\circ$ \\ 
%8 & /$45^\circ$ & $135^\circ$ \\
8 / 8 & $0^\circ$ / $45^\circ$ & $135^\circ$ / $180^\circ$ & $\measuredangle = 135^\circ$ \\
%8 & $45^\circ$ & $180^\circ$ \\
\bottomrule
\end{tabular}

\vspace{-3mm}
\end{table}

%\begin{table}
%\caption{Description of evaluation dataset.}
%\label{tab:evalset}
%\begin{tabular}{cccc}
%\toprule
%& \multicolumn{2}{c}{Movement Type} & \\ 
%\# Samples & Source 1 & Source 2 & Details \\ 
%\midrule
%16 & CW & CW & $\measuredangle > 45^\circ$ \\
%16 & CCW & CCW & $\measuredangle > 45^\circ$ \\
%16 & CW & CCW & Sources intersect \\
%\bottomrule
%\end{tabular}
%\end{table}

%\begin{table}
%\caption{Description of evaluation dataset.}
%\label{tab:evalset}
%\begin{tabular}{cccc}
%\toprule
%& \multicolumn{2}{c}{Movement Type} & \\ 
%\# Samples & Source 1 & Source 2 & Details \\ 
%\midrule
%16 & CW & CW & $\measuredangle > 45^\circ$ \\
%16 & CCW & CCW & $\measuredangle > 45^\circ$ \\
%16 & CW & CCW & Sources can cross \\
%\bottomrule
%\end{tabular}
%\end{table}

\subsection{Dataset with stationary sources}
\label{sec:stationary}
In order to compare the performance of the proposed algorithm against conventional methods assuming stationary sources \cite{sawadaCNMFtaslp,nikunenCNMFtaslp,nikunenCNMFicassp}, we include additional evaluation dataset with completely stationary sources. We use the dataset introduced in \cite{nikunenCNMFtaslp} consisting of two simultaneous sound sources. In short the dataset contains speech, music and noise sources convolved with RIRs from various angles captured in a regular room (7.95 m x 4.90 m x 3.25 m ) with a reverberation time $T_{60} = 350$ ms. The array used for recording is exactly the same as the one used in the datasets introduced in Section \ref{sec:datasets} and more details of the recordings can be found from \cite{nikunenCNMFtaslp}. In total the dataset contains 48 samples with 8 different source types and 6 different DOA combinations. Each sample is 10 seconds in duration. The different conditions are summarized in last tabular of Table \ref{tab:dataset}. 

%\begin{table}
%\setlength{\tabcolsep}{4pt}
%\caption{Description of stationary sources dataset.}
%\label{tab:dataset}
%\begin{tabular}{cccc}
%\multicolumn{4}{l}{\textbf{Dataset from \cite{nikunenCNMFtaslp}} \\
%\toprule
%Number of & & & \\
%samples & Source 1 & Source 2 & Details \\ 
%\midrule
%8 & CW & CW & $\measuredangle > 45^\circ$ \\
%8 & CCW & CCW & $\measuredangle > 45^\circ$ \\
%8 & CW & CCW & Sources intersect \\ 
%8 & CW/CCW & S (babble) & SNR = -5, -10 and -15 dB \\
%8 & CW/CCW & S (music) & SNR = -10 dB \\
%8 & CW/CCW & S (music) & SNR = -10 dB \\
%\bottomrule
%& & & \\
%\end{tabular}
%\end{table}

\subsection{Experimental setup}
\label{sec:procparam}
For the WGMM parameter estimation the peaks in the SRP function were enhanced by exponentiation $\zoj^{(3/2)}$, which emphasizes high energy peaks (direct path) and low energy reflected content is decreased. This was found to improve operation in moderate reverberation. A five-component ($\Gmax = 5$) WGMM model \eqref{eq:WGMMpdf} was estimated from the SRP and parameters from a previous frame were used as an initialization for next frame. % (first frame was initialized randomly). 
The criteria for removing WGMM measurements were set to values of $\Zsigma > 0.6 \; \mathrm{rad} \; (34^\circ)$ and $\Za < 0.15$ by visually inspecting the development set results.

%Due to the randomization in the particle filtering resulting to slightly different results in every algorithm run, the study of effect of acoustic tracker parameters was also done experimentally using subset of the test signals and visually inspecting results (source detection, prediction over inactive segments, etc.).

The acoustic tracker parameters were optimized by maximizing the development set tracking performance. The parameters were set to the following values. Average variance of the WGMM measurements was scaled to $\sigma^2 = 0.25$ for each processed signal to be in appropriate range for the particle filtering toolbox\footnote{\url{http://becs.aalto.fi/en/research/bayes/rbmcda/}}. The clutter prior probability was fixed for all measurements to $CP = 0.1$. In the particle filtering framework the life time of the target is modeled using a gamma distribution with parameters $\alpha$ and $\beta$. The best tracking performance was achieved with $\alpha = 3$ and $\beta = 4$. The target initial state was fixed to $\state_\jind^{(\qind)} = [\cos(\pi), \sin(\pi), 0.1, 0.1]$. % with higher uncertainty given to the velocities. 
The pre-set prior probability of source birth was set to $BP = 0.005$.

The parameters of the multichannel NMF algorithm were set as follows: the window length was 2048 samples with 50\%  overlap and 80 NMF components were used for modeling the magnitude spectrogram. The entire signals were processed as whole. %Notes regarding block-wise processing with the proposed model can be found from discussion Section \ref{sec:discussion}. 
Before restoring the spatial weights by Equation \eqref{eq:Zrestore} a minimum and maximum variance for $\sigma_{\jind,\qind}$ were set to 0.025 and 0.3, respectively. This was done in order to avoid unnecessarily wide or narrow spatial window (as can be seen from Figure \ref{fig:Background}). With the chosen minimum and maximum values for $\Zsigmap$, the constant in Equation \eqref{eq:Zrestore} becomes $c = \max(\sigma_{\jind,\qind}^2) + \min(\sigma_{\jind,\qind}^2) = 0.325$. The background source threshold for setting the spatial weights active was set to $0.01$, which corresponds to approximately $\pm30^\circ$ exclusive spatial window for the actual tracked sources when the tracker output state variance is at its minimum %$(0.025) \rightarrow -\sigma_{\jind,\qind}^2 + c = 0.3$ 
indicating high certainty of source being present at the predicted direction.

%In the first CNMF stage the algorithm was run for 50 iterations starting from parameters initialized by random values drawn from uniform distribution. 
%The spatial weights were initialized by SRP calculated as specified in Chapter 5 Equation (6.23) of \cite{tashevDOAkirja}. %The first CNMF stage was parameterized for $\qmax = 2$. After the acoustic tracking the dimensions of the NMF component to source association $\bkp$ were adjusted according to the number of actual found sources, added rows were initialized with random parameters. 
%The total number of separated source signals $\qmax$ is the number of sources detected by the acoustic multi-tracking algorithm added with the background source/track. 

%For restoring the spatial weights based on the DOA trajectories (Equation \eqref{eq:Zrestore}) the fixed variance of $\sigma^2 = 0.2$ was used. 

\subsection{Acoustic tracking performance}
\label{sec:akuperf}
The acoustic tracking performance is evaluated against the hand-annotated ground truth source trajectories by using the accuracy (mean absolute error) and recall rate as the metrics. The tracking error for each source in each time frame with $2\pi$ ambiguity is specified as
\begin{equation}
e_{\jind,\qind} = \Zanno - \Zmup = \tilde{e}_{\jind,\qind} + 2 \pi N, \quad N \in \mathcal{Z}
\end{equation}
where $\Zanno$ denotes the annotated DOA of $\qind$th source in time frame $\jind$ and $\Zmup$ is obtained using Equation \eqref{eq:PFtoAngle}. Using the error term $\tilde{e}$ which is wrapped to $[-\pi, \pi]$, we specify mean-absolute error ($\mathrm{MAE}$) as 
\begin{equation}
\mathrm{MAE} = \sum_{\qind=1}^{\hat{P}} \frac{1}{\jmax} \sum_{\jind=1}^{\jmax} |\tilde{e}_{\jind,\qind}|,
\end{equation}
where $\hat{P}$ is the number of annotated sources. 

The recall rate is defined as the proportion of time instances the detected source is correctly active with respect to when the source was truly active and emitting sound. The ground truth of the active time instances is obtained by voice activity detection (VAD) using the close-field signal of the source. The VAD is used in order to take into account that some utterances start 1 to 2 seconds after the beginning of the signal even though the annotations are continuous for the whole duration of recordings. Additionally, if the tracked source dies before the end of the signal during a pause of speech, the duration of the pause is not accounted for as a recall error, but the remaining missing part is. We will denote the recall rate using variable $\mathrm{recall} \in [0,1]$.

The proposed method uses multi-target tracker that can detect arbitrary number of sources and trajectories denoted as $\qmax$. For evaluation of tracking and separation performance we need to match the annotated sources $1,...,\hat{P}$ and detected sources $1,...,\qmax$ by searching trough all possible permutations $r$ of the detected sources denoted as $\mathbf{P}_r : \{1,...,\qmax\} \rightarrow \{1,...,\hat{P}\}$. The permutation matrix $\mathbf{P}_r$ is applied to change the ordering in which detected sources are evaluated against the annotations.

%$r = 1,\dots,\mathsf{R}$, where $\mathsf{R} = \frac{!\qmax}{!(\qmax-\hat{P})}$ 

We propose to choose the permutation $r$ for final scoring that maximizes combination of $\mathrm{MAE}$ and $\mathrm{recall}$. First $\mathrm{MAE}$ is converted into a proportional measure $\mathrm{MAER} = 1-( \mathrm{MAE} / \pi) \in [0,1]$, where 1 denotes zero absolute error and 0 denotes maximum $\pi \; \mathrm{rad} = 180^\circ$ tracking error at all times. Summing the $\mathrm{MAER}$ and the recall rate with permutation $r$ applied to the estimated sources equals to
\begin{equation}
\label{eq:Fscore}
\mathrm{F}_{r} = \mathrm{MAER}_{r} + \mathrm{recall}_{r},
\end{equation}
which is referred to as overall accuracy. The best permutation for each signal is chosen by finding the minimum value of $\mathrm{F}_{r}$ over all permutations indexed by $r$. The combination of both measures is used to avoid favoring permutations with very short detected trajectories with small $\mathrm{MAE}$ over longer trajectories with slightly larger $\mathrm{MAE}$, for example accurate tracking of a single word from the entire utterance. Additionally, we do not consider and compensate for cases where one sound source is correctly tracked by two trajectories with discontinuity during the pauses in speech. The effect of this is negligible due to the short test signals used (10 seconds). % and speech in the recorded signals being continuous with only natural pauses between words and sentences.

% andthis permutation is used in evaluation of overall MAE, recall rate and separation performance. 

The acoustic tracking performance averaged over all signals in all datasets and source detection performance is reported in Table \ref{tab:trackeval}. The tracking error measured by $\mathrm{MAE}$ is below 10 degrees for datasets with two sources, which can be regarded as a good result. Noticeably the accuracy of the tracking is even better with the evaluation dataset. However the recall rate drops by 4\% mostly due to late detection of sources, which displays the difficulty of setting the optimal values for parameters controlling the birth and death of sources in the particle filtering. In general, a low recall rate can be considered as conservative and only detecting and tracking dominant portions of sources. Alternatively, optimizing the parameters for 100\% recall rate would lead to detection of numerous phantom sources, caused by reverberation and noise in the recordings. The recall rate and tracking accuracy are noticeably decreased for the evaluation dataset with three simultaneous sources.

As indicated by the second chart in Table \ref{tab:trackeval}, the percentage of correctly detected number of sources is approximately 80\% for both two source datasets and drops down to 56\% for the dataset with three sources. The errors in source detection are mostly caused by overdetection in the case of two simultaneous sources, whereas in the more difficult scenario of three sources the underdetection is also a significant cause of error. %The false source detections may be either due to reflections or simply by noise. 
%Detection of false sources  close to the actual correct trajectory may deteriorate the separation performance due to separation mask estimation parameterized by the DOA.

\begin{table}
\caption{Acoustic tracking results.}
\label{tab:trackeval}
\centering
\begin{tabular}{cccc}
\multicolumn{4}{l}{\textbf{Tracking performance}} \\
\toprule
%& \multicolumn{2}{c}{Movement Type} & \\ 
& \textbf{Dev.} & \textbf{Eval.}  & \textbf{Eval.} \\ 
\textbf{Criteria} & \textbf{(2 sources)} & \textbf{(2 sources)}  & \textbf{(3 sources)} \\ 
\midrule
$\mathrm{MAE}$ & $~7.3^\circ$ &  $~6.1^\circ$ & $10.5^\circ$ \\
$\mathrm{recall}$ & 86.3\% & 82.2\% & 64.7\% \\
\bottomrule
& & & \\
\end{tabular}

\begin{tabular}{cccc}
\multicolumn{4}{l}{\textbf{Source detection performance}} \\
\toprule
%& \multicolumn{2}{c}{Movement Type} & \\ 
& \textbf{Dev.} & \textbf{Eval.}  & \textbf{Eval.} \\ 
\textbf{Criteria} & \textbf{(2 sources)} & \textbf{(2 sources)}  & \textbf{(3 sources)} \\ 
\midrule
$\qmax == \hat{P} $ & 79.2\% & 81.3\% &  50.0\% \\
$\qmax > \hat{P} $ & 20.8\% & 16.7\% & 25.0\% \\
$\qmax < \hat{P} $ & ~0.0\% & ~2.0\% & 25.0\% \\
\bottomrule
\end{tabular}
\vspace{-3mm}
\end{table}

%The acoustic tracking performance was measured using mean absolute error (MAE) and recall rate. 

%The details of the acoustic tracking evaluation is explained in page 16 of project meeting slides \url{nokia121114.pdf}. %The final acoustic tracking results can be found from page 30 of project meeting slides \url{nokia111214.pdf}. 
%The best performing realization achieved MAE of 11.8 azimuthal degrees on average while having a recall rate of 83\%. The imperfect recall rate is mainly caused by delayed detection of the sources in the tracking stage. This is due to the tracking realization being online while the overall framework is offline. 

\subsection{Source separation performance}
\label{sec:SSp}

%\subsection{Evaluation procedure and contrast methods}
\subsubsection{Separation evaluation criteria}
We evaluate the separation performance of the proposed algorithm using the following objective separation criteria with the close-field microphone signal as a reference. From the separation evaluation toolbox proposed in \cite{sisec} we have included signal-to-distortion ratio ({SDR}) and signal-to-interference ratio ({SIR}) evaluated in short segments of 200 ms. The score of each segment in dB scale is converted to linear scale and averaged over all segments after which the resulting average is converted back to dB scale. The resulting metrics are abbreviated as segmental SDR ({SSDR}) and segmental SIR ({SSIR}). %  and the range of values was limited to be from $-5$ dB to $+20$ dB for SDR and SIR in each segment before averaging. 
The use of segmental evaluation was chosen due to the operation of BSSeval, which projects the separated signal into reference signal subspace and assumes that this projection is stationary. However, in the case of moving sound sources and reference by close-field microphone the initial delay to the far-field array and the room reflections change from frame to frame which requires the projection operator to be also time-variant. This is achieved by assuming projection stationarity within each 200ms segment. %Evaluating the separation performance in on segment (entire signal at once) with ideal ratio mask (IRM) separation applied to the mixture achieves an average SDR below 0dB which does not reflect the actual achieved performance.
%The use of short segments was chosen to alleviate the effect % of evaluation against momentary full noisy segment or near silence,
% of recall errors, \ie, evaluation of zero signal against the ground truth.  %The SDR and SIR scores of each segment are averaged in linear domain and brought back to decibel scale, which reduces the effect of single small or large separation score in averaging. 
%Additionally, a voice activity detection (VAD) is used to determine which of the reference signal segments contains actual speech and only those segments where VAD is active are included in the evaluation. The details of the VAD can be found from page 15 at project meeting slide \url{nokia121114.pdf}. 
Other metrics include the short-time objective intelligibility measure ({STOI}) \cite{STOI} which is used to predict the intelligibility of the separated speech in comparison to the reference signal, and the frequency-weighted segmental signal-to-noise ratio ({fwSegSNR}) \cite{fwSNR}. The latter metrics, STOI and fwSeqSNR, were calculated without segmenting.
% and overall perceptual score (OPS) from \cite{OPSscore}. 
%Evaluation criteria for acoustic tracking peformance are introduced in Section \ref{sec:akuperf}

\subsubsection{Reference methods}
The description of methods whose separation performance is evaluated are given
%\begin{enumerate}
%\item mic : Microphone signal from the array (ch \#1).
%\item DSB : Delay-and-sum beamforming.
%\item MVDR : Minimum variance distortionless beamforming.
%\item CNMF : Proposed method.
%\item CNMF ann. : Proposed method with ground truth annotations as source trajectories.
%\item IRM : Ideal ratio mask separation.
%\item CNMF s. : CNMF-based separation assuming stationary sources \cite{nikunenCNMFicassp}.
%\end{enumerate}
in Table \ref{tab:methods} and can be summarized as follows. The plain microphone signal from the array acts as a lowest performance baseline whereas the \textit{IRM} indicates an upper limit. The tracking information is also used in the beamforming (\textit{DSB} and \textit{MVDR}) to specify the weights $\mathbf{w}_{\iind\jind,\qind}^H$ to enhance the signal at each time frame originating from the estimated DOA.
%thus the acoustic tracking realization was only run once and the resulting DOA trajectories were shared between proposed and beamforming methods. 
%This was done to rule out any difference caused by slightly different tracking result due to randomization in the particle filtering. 
The separation performance of proposed method was also evaluated using the ground truth DOA trajectories for specifying the source movement for the multichannel NMF part. This evaluation indicates the highest achievable separation performance with perfect tracking information and the results can be also used for validating the robustness of the overall proposed separation algorithm towards small tracking errors.

The \textit{MVDR} beamforming was implemented using the sample covariance method \cite{tashevDOAkirja} for estimating the noise covariance matrix. The noise covariance was estimated from $M = 20$ previous frames with respect to each processed frame and it captures the stationary noise statistics as well as the immediate spectral details of interfering speech sources. Additionally, a diagonal loading ($\sigma = 5$) of noise covariance matrices was applied to improve robustness of the \textit{MVDR} beamformer. These parameters were optimized using the development set.  %As based on assumption of stationary sound sources and estimation of separation masks by complex-valued NMF with time-invariant SCMs.

\begin{table}
\caption{Description of compared separation methods.}
\label{tab:methods}
\centering
\begin{tabular}{{p{1.2cm}p{6.2cm}}}
\toprule
\multicolumn{1}{c}{\textbf{Abbrv.}} & \textbf{Description} \\ 
\midrule
\multicolumn{1}{c}{\textit{mic}} & Microphone signal from the array (ch \#1). \\
\multicolumn{1}{c}{\textit{DSB}} & Delay-and-sum beamforming. \\
\multicolumn{1}{c}{\textit{MVDR}} & Minimum variance distortionless beamforming. \\
\multicolumn{1}{c}{\textit{MNMF sta.}} & Multichannel NMF assuming stationary sources \cite{nikunenCNMFicassp}. \\
\multicolumn{1}{c}{\textit{MNMF}} & Proposed method, i.e. multichannel NMF with time-varying SCM model. \\
\multicolumn{1}{c}{\textit{MNMF ann.}} & Proposed method with ground truth annotations as source trajectories. \\
\multicolumn{1}{c}{\textit{IRM}} & Ideal ratio mask separation. \\
\bottomrule
\end{tabular}
\vspace{-3mm}
\end{table}

\begin{figure*}[tb]
  \centering
  \includegraphics[width = 1.0\linewidth]{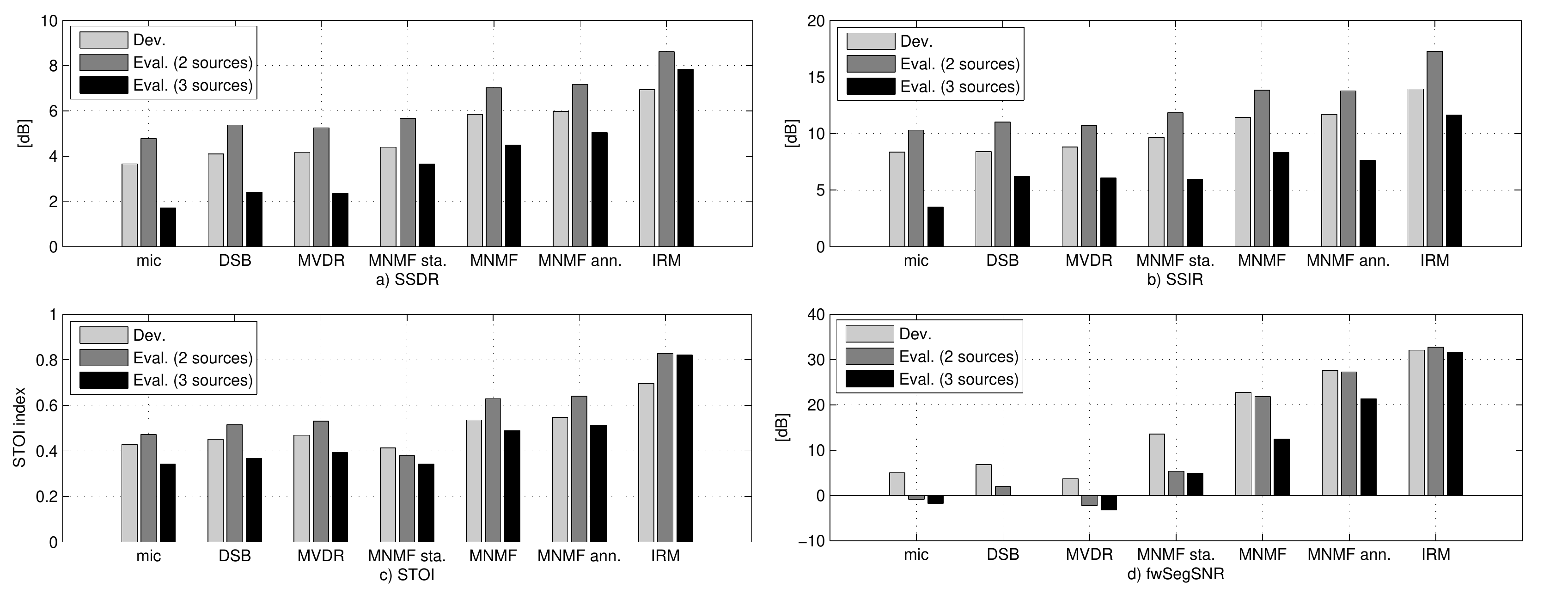}
	\caption{Separation performance measured using various objective separation criteria: SSDR, SSIR, STOI, fwSegSNR.}
  \label{fig:results1}
	\vspace{-3mm}
\end{figure*}

\subsubsection{Separation results}
The separation performance measured by SSDR, SSIR, STOI and fwSegSNR is calculated with the source permutation obtained by minimizing the accuracy criterion specified in Equation \eqref{eq:Fscore} and the results are averaged over all sources and mixtures. The separation performance for all the tested methods with all the considered criteria are given in Figure \ref{fig:results1} (a)-(d). 

%\subsubsubsection{Baseline performance analysis}
Evaluation with the mixture signal (mic) indicates a baseline performance resulting to SSDR of approximately 4~dB for two simultaneous sources and 2~dB for three sources. Such high absolute performance for mixture signal is because of the segmental evaluation. In contrast, evaluating the entire signals in one segment resulted into negative average SDR for all tested methods (from -6 dB SDR for the mixture to -1 dB for the IRM) while the relative differences remained the same as reported in the Figure \ref{fig:results1}. The absolute results obtained this way did not reflect the subjective separation performance and are not reported in the paper. The overall low scores were assumed to be caused by the problems of projection operation in BSSeval toolkit, discussed in the beginning of Section \ref{sec:SSp}.

The beamforming methods, \textit{DSB} and \textit{MVDR}, consistently improve SSDR, SSIR and STOI in comparison to microphone signal. However, the overall improvement in all datasets is relatively poor: SSDR improvement varies from 0.45~dB to 0.70~dB, and STOI barely reaches the index of 0.5, indicating low predicted intelligibility for the separated sources. Additionally, \textit{MVDR} beamforming has negative effect on the fwSegSNR criterion, which may be caused by unwanted canceling of target source due to the rudimentary noise covariance estimation method employed. \textit{DSB} on the other hand does not have this negative effect. With all other evaluated criteria \textit{MVDR} beamforming exceeds the \textit{DSB} performance with a small margin.

When violating the moving sources assumption and using multichannel NMF aimed for separation of stationary sound sources \cite{nikunenCNMFicassp}, the performance is poor especially with STOI that decreases below the array microphone baseline. The other evaluated metrics are similar to beamforming approaches. When the source is momentarily at the target static direction (blindly estimated in \cite{nikunenCNMFicassp}) the separation quality is high, but as the source moves it shifts out of spatial focus of the separation. %The development dataset contains signals consisting of a stationary source along with one moving source and successful estimation the stationary source in noise free scenario leads to also fair separation of the moving source due to the Wiener filter-based reconstruction of signals. This phenomenon is arguably visible from the results of the stationary CNMF which are noticeably better for the evaluation dataset in comparison to the results with other datasets only consisting of moving sound sources.

%\subsubsubsection{Proposed method performance analysis}
The results regarding the proposed method can be summarized by stating that it significantly increases the separation performance over the beamforming methods \textit{DSB} and \textit{MVDR}: in case of two sources the SSDR improves approximately by 1.5~dB and improvement is even greater for three simultaneous sources (2~dB). SSIR improvement follows the trend of SSDR and STOI increases approximately by an index of 0.1. The use of ground truth annotations in the case of two simultaneous sources does not significantly increase the performance: SSDR increases only by 0.13~dB and 0.15~dB, SSIR is unchanged and STOI increases by 0.01. This validates the good acoustic tracking results reported in Table \ref{tab:trackeval}. The separation performance with the three sources follows the poorer tracking performance and the use of annotations has greater impact on improving the objective criteria, SSDR is increased by 0.5~dB by use of annotations but interestingly the SSIR decreases. This may be due to the annotations being always active even if the speech starts 1 to 2 seconds from the beginning of the signal resulting to nonzero separation output for the annotations, whereas in the actual tracking implementation the source signal is truly zero until it is detected.

The behavior of all methods with all criteria is consistent with all datasets. Although the absolute differences in objective intelligibility are small, the increase in STOI by the proposed method over the beamforming methods is greater than the STOI improvement by beamforming over the  microphone signal. The overall difficulty of each dataset can be estimated from the IRM performance, which indicates that the actual evaluation dataset with two sources is less difficult than the development dataset. %This can be explained by not having the recordings of stationary sources with very reverberant propagation paths in the evaluation dataset and such sources cannot be accurately reconstructed by only using time-frequency masking. Their highly reflected propagation paths are also problematic for acoustic tracking and separation mask estimation parameterized by the DOA.
The performance gap between the proposed method and \textit{IRM} separation is considerable especially in case of three simultaneous sources. This is due to the fact that \textit{IRM} is not much affected by the adding of third source, since speech is relatively sparse in time frequency domain and good separation can be achieved with oracle masks even with three simultaneous speakers. Evaluation of IRM performance by SSDR and SSIR for two source dataset indicates not as big difference in comparison to the proposed method. However, the IRM performance is also limited by the fact that far-field and close-field signal spaces are extremely different and time-frequency masking cannot recover the close-field signal perfectly due to mixture phase is used. In subjective evaluation \textit{IRM} preserves the intelligibility of the speech much better than any separation method which is also indicated by the good results in the objective evaluation of intelligibility, i.e. STOI for \textit{IRM} is around 0.7-0.8.

%SDR is approximately doubled with the use of IRM in comparison to the proposed method and IRM also results in significant increase in SIR and almost 0.2 increase in STOI with respect to the proposed method. The modest overall scores obtained with the oracle IRM separation ( 0.7-0.8 STOI index) indicate that separating sound sources with time-varying mixing properties even in moderate reverberation is a difficult task. 

As a final result we provide a comparison of separation performance obtained with similar DOA-based spatial and spectrogram factorization models assuming stationary sources \cite{sawadaCNMFtaslp,nikunenCNMFtaslp,nikunenCNMFicassp}. The evaluation dataset consist of all sources being stationary, see Section \ref{sec:stationary}. The proposed algorithm was run as is with the exception that source reconstruction was done without DSB, since the reference signals are reverberated source signals and evaluation in \cite{nikunenCNMFicassp} is based on the spatial images of the sources \cite{sisec2}. The details of evaluation procedure and reference results are as presented in \cite{nikunenCNMFicassp}. In theory, if the source DOA trajectory estimation would be perfect, similar results between multichannnel NMF-based methods regardless of source movement assumption should be obtained. However, methods proposed in \cite{nikunenCNMFtaslp,nikunenCNMFicassp} also update elements of the DOA kernels (Equation \eqref{eq:Wkernel}) whereas in this work they are fixed to analytic anechoic array responses.

The SDR, SIR, SAR and ISR are given in Table \ref{tab:results2}, which shows that the SDR performance of the proposed method is lower in comparison to methods utilizing the stationary assumption, while SIR is highest among the tested methods. %It is unclear if the difference in performance is only caused by not updating the elements of the DOA kernels or a combination of differences in the algorithms, for example the spatial weights being time dependent and affected by small errors in DOA trajectory estimation making the method deviate from the 
The average tracking error (MAE) for the dataset was $8.8^\circ$ and recall rate was 79\% which are similar to the tracking performance for other 2 source datasets given in Table \ref{tab:trackeval}. The comparison to multichannel NMF models assuming stationary source motivates the future work for reducing the performance gap while assuming moving sound sources and possible research directions are discussed in the next section.

\begin{table}[t]
\centering
\begin{tabular}{|c|cccc|}
  \hline
  \textbf{Method} & \textbf{SDR} & \textbf{SIR} & \textbf{SAR} & \textbf{ISR} \\ 
   \hline
  MNMF proposed & 3.2 dB & \textbf{9.0} dB & 7.0 dB & 5.8 dB \\
	MNMF sta. \cite{nikunenCNMFicassp} & \textbf{5.6} dB  & 6.8 dB & \textbf{13.1} dB & 9.9 dB \\
  MNMF sta. \cite{nikunenCNMFtaslp} & 4.8 dB & 8.1 dB & 10.3 dB & \textbf{10.5} dB \\
  MNMF sta. \cite{sawadaCNMFwaspaa} & 3.7 dB & 4.5 dB & 12.7 dB & 8.4 dB \\
  ICA \cite{SawadaICAdoa} & 2.0 dB & 4.5 dB & 8.2 dB & 6.9 dB \\
  \hline
\end{tabular}
\caption{Results of dataset and methods from \cite{nikunenCNMFtaslp} and \cite{sawadaCNMFwaspaa} with two simultaneous stationary sources.}
\label{tab:results2}
\vspace{-4mm}
\end{table}

%The results indicate that the best performing acoustic tracker realization also translates to best separation performance in terms of both SSDR and STOI. 
%The objective separation and predicted intelligibility are increased from the 2014 realization. The informal listening of the results indicates even bigger separation peformance improvement than the objective measures. 

% A different strategies for improving the perceptual quality of the separated signals by smoothing the Wiener-masks is documented in project meeting slides \url{nokia111214.pdf} and the objective separation results are given in page 32. In conclusion, by time-averaging the NMF gain responses improves both the objective separation but also the perceived quality of the separated signals, but slightly decreases the STOI.

\section{Discussion}
\label{sec:discussion}
In this section we present a few remarks regarding the algorithm development choices and possible future work for improving and extending the method. 
 
The strategy of using the estimated source DOA trajectories for definition of the SCM model in \eqref{eq:CNMFmodel2} means effectively %defining the source SCMs by 
using only channel-wise time differences and assuming anechoic environment. This strategy can be questioned in comparison to also updating the channel-wise level differences as in \cite{nikunenCNMFtaslp,nikunenCNMFicassp}. However, the difficulty of updating the level differences between input channels lies within the fact that with moving sources there may be only very few frames of data observed from each direction and investigation of the updating $\Wmat$ in such setting was left for future work. 

The multichannel NMF model \eqref{eq:CNMFmodel2} would allow to use multichannel Wiener filter (MWF) for source reconstruction as in \cite{sawadaCNMFtaslp}. Informal experiments showed inferior performance with MWF in comparison to chosen combination of single-channel Wiener filter and DSB. There are several possible reasons to explain the findings. The multichannel model used for representing source SCMs relies only on the anechoic responses which can be suboptimal for constructing the MVF for source reconstruction. Additionally, errors in source SCM estimation can lead to unexpectedly sharp spectral and spatial responses for source reconstruction with MVF. The strategy of single-channel Wiener filter and DSB is argued to be less destructive with respect small estimation errors. %Our results are in agreement with the findings mentioned in \cite{sawadaCNMFtaslp} which report that empirically better results were obtained with single-channel Wiener filter in case of Euclidean based cost function.

%Additionally, we did not consider processing of long and continuous signals in the work, but block-wise operation of the CNMF model parameter estimation can be considered as a simple extensions to the presented work. The proposed model can be thought to be segmentally applied to short overlapping segments of 2 to 3 seconds of data and maintaining the source association between segments by the source trajectory information. Inheritance of learned spectral basis can be used to speed up the converge with each new segment of data.

%The novelty of the proposed method can be summarized as extending the parameterization of SCMs in complex-valued NMF to be time-variant and combining the source tracking information with such model. The use of source tracking information then allows estimation of separation masks corresponding the spatial evidence emitted from the estimated directions in each time frame. Even though the work does not claim any novelty in the field of acoustic tracking, 

Analysis of the tracking performance indicated that fairly accurate source DOA trajectories can be estimated with existing methods in realistic capturing conditions which justifies the applicability of the proposed separation algorithm for general use. %Therefore some tracking performance discussion related to operation realistic environments are presented next.
Tracking errors may be caused by erroneously representing a single source with two consecutive but separate tracks due to pauses in speech. Also in the case of intersecting DOA trajectories, the estimated tracks can switch the actual acoustic targets, \ie, source 1 continues to track acoustic evidence of source 2 and vice versa. It should be noted that we did not account for the above problems in the tracking and separation performance evaluation. % and to alleviate effect of possible splits and switches.

%The rotating vector model for the state space model used in the paper is not theoretically optimal in a sense that the observations and the underlying state are truly 1D quantities. The conversion from 1D to 2D according to Traa in \cite{traaWrapped} increases the effect of noise in the system to be tracked. The use of 1D state and 2D observation was investigated during the algorithm development but resulting non-linear state space model and using EKF to update the particle distributions resulted to worse tracking performance.
%However, it requires using non-linear transition matrix for the state space model and the results were worse than linear dynamic model with standard KF for particle updates.. % and the statistics of the particles needs to be updated using EKF. 
%The robustness and the tracking performance of the non-linear approximation by EKF were subpar in comparison to linear dynamic model with standard KF for particle updates. 
%As a future work for improving the proposed algorithm, 
%The problem of tracking wrapped quantities using 1D state could be addressed via wrapped Kalman filtering as proposed in \cite{traaWrapped}.

The extremely good results of using of deep learning for speech separation \cite{DNN1,DNN2,deepClust} are quickly replacing the use factorization based models in source separation. With multichannel audio the spatial parameters being complex-valued require use of other approaches for SCM estimation, for example in \cite{DNNcovar2rev} DNNs are used for spectrogram estimation while SCMs are estimated using a probabilistic model and EM-algorithm. %The question of how quickly deep learning approaches will be adopted for multichannel source separation and how big advances are gained is yet to be seen. 
The strength of the proposed method compared to DNN-based separation is that it operates on spatial information and spectral factorization of the observed data only and works relatively well in any scenario and all sound content (music, noise, everyday sounds) without any training material. 

%Furthermore, this causes the representation of source magnitude spectrum $\sFDhat$ to become unreasonable using a NMF model, which utilizes the long term redundancy and recurrent spectral shapes in representing the spectrogram. Additionally, the estimate of spatial parameters from a one STFT frame is extremely unreliable and prone to noise and clutter in the spatial evidence (the phase difference). If a source is momentarily inactive, there is no spatial evidence of it, and usually in case of overlapping sound sources the most energetic source may dominate the spatial features in respective time frame. 

\section{Conclusions}
\label{sec:Conclusion}
In this article a separation method for moving sound sources based on acoustic tracking and separation mask estimation by multichannel non-negative matrix factorization was proposed. We analyzed the objective separation performance and the proposed method exceeded the conventional beamforming using the same tracking information by a fair margin. The comparison against ground truth source DOA trajectories indicated only minor impairment to objective separation performance. Additionally, analysis of the acoustic tracking realization showed good performance, recall rate over 80~\% and absolute tracking error less than 10 degrees with two simultaneous moving sound sources. In conclusion the proposed method was shown to be robust and capable of separating at least two moving targets from mixtures recorded with a compact sized microphone array in realistic capturing conditions.

%\clearpage
%\vfill
%\pagebreak

\bibliographystyle{IEEEtran}
\bibliography{refs}

\end{document}